\documentclass[epsfig,preprint]{aastex}
\begin{document}

\newcommand{\hi}{$h^{-1}$~}
\newcommand{\kms}{~km~s$^{-1}$}

\title{The Galaxy Populations of X-Ray Detected, Poor Groups}

\author{Kim-Vy H. Tran } 
\affil{Department of Astronomy \& Astrophysics, University of
California, Santa Cruz, CA 95064} 
\email{vy@ucolick.org} 

\author{Luc Simard \altaffilmark{1} }
\affil{University of California Observatories/Lick Observatory, 
University of California, Santa~Cruz, CA 95064}
\email{lsimard@as.arizona.edu}

\author{Ann I. Zabludoff \altaffilmark{1}}
\affil{University of California Observatories/Lick Observatory, 
University of California, Santa Cruz, CA 95064}
\email{azabludoff@as.arizona.edu}

\author{John S. Mulchaey }
\affil{Observatories of the Carnegie Institution of Washington, 813 
Santa Barbara St., Pasadena, CA 91101}
\email{mulchaey@ociw.edu}

\altaffiltext{1}{Department of Astronomy and Steward Observatory, University of Arizona, 
Tucson, AZ 85721}

\begin{abstract}

We determine the quantitative morphology and star formation properties
of galaxies in six nearby X-ray detected, poor groups using
multi-object spectroscopy and wide-field $R$ imaging.  The mean
recessional velocities of the galaxy groups range from 2843 to 7558
\kms.  Each group has 15 to 38 confirmed members ranging in luminosity
from dwarfs to giants ($-13.7\geq M_R-5$log$h\geq -21.9$).  We measure
structural parameters for each galaxy by fitting a PSF-convolved, two
component model to their surface brightness profiles.  To compare
directly the samples, we fade, smooth, and rebin each galaxy image so
that we effectively observe each galaxy at the same redshift
(9000\kms) and physical resolution (0.87\hi kpc).  The structural
parameters are combined with [OII] measurements to test for
correlations between morphological characteristics and current star
formation in these galaxies.  We compare results for the groups to a
sample of field galaxies.  We find that: 1) Galaxies spanning a wide
range in morphological type and luminosity are well-fit by a de
Vaucouleurs bulge with exponential disk profile.  2) Morphologically
classifying these nearby group galaxies by their bulge fraction
($B/T$) is fairly robust on average, even when their redshift has
increased by up to a factor of four and the effective resolution of
the images is degraded by up to a factor of five.  3) The fraction of
bulge-dominated systems in these groups is higher than in the field
($\sim 50$\% vs. $\sim20$\%). 4) The fraction of bulge-dominated
systems in groups decreases with increasing radius, similar to the
morphology-radius ($\sim$ density) relation observed in galaxy
clusters.  5) Current star formation in group galaxies is correlated
with significant morphological asymmetry for disk-dominated systems
($B/T<0.4$).  6) The group galaxies that are most disk-dominated
($B/T<0.2$) are less star forming and asymmetric on average than their
counterparts in the field.

\end{abstract}


\section{Introduction}

How the evolution of galaxies is affected by environment continues to
be a fundamental question in astronomy.  Recent work by Schade
$et~al.$ (1995, 1996), Abraham $et~al.$ (1996a, 1996b), Ratnatunga
$et~al.$ (1999), and Simard $et~al.$ (1999, 2000) has focused on
quantifying the morphologies of galaxies and relating them to
environment and spectral star formation indicators.  These efforts
concentrate on field or cluster galaxies.  Little has been done to
quantitatively characterize galaxies in the intermediate, and common,
environment of poor groups where the factors that might influence
galaxy evolution differ from those in clusters or the rarefied field.
One puzzle is that some poor groups with extended X-ray emission share
certain properties with rich clusters, such as the shape of the galaxy
luminosity function (Zabludoff \& Mulchaey 2000) and a high early-type
galaxy fraction (Zabludoff \& Mulchaey 1998), but have lower velocity
dispersions and galaxy number densities.  By fitting 2D surface
brightness profiles and examining the residual light after the model
is subtracted, it now is possible to derive structural parameters for
a large sample of group galaxies that may then be compared with
results from other environments and with the galaxies' spectral
properties.  In particular, we can quantitatively and uniformly type
galaxies by their bulge fraction, determine the fraction of galaxies
with significant asymmetry, measure whether groups have a
morphology-radius ($\sim$ density) relation like clusters, compare
group galaxy morphologies with the field, and assess whether the
current star formation level in group galaxies is different than that
in field galaxies of the same type.

As group galaxy number densities are closer to that of the field than
of rich clusters, do the group galaxy populations differ significantly
from the field?  Early-type galaxies are the minority in the field,
making up only $\sim$23-29\% of the field population (van den Bergh
$et~al.$ 1996; Driver $et~al.$ 1995; Shapley \& Ames 1932).  Given the
morphology-density relation and the low galaxy number densities and
velocity dispersions of groups, it is surprising that high early-type
fractions consistent with those of rich clusters have been found in
some X-ray detected groups (Mulchaey \& Zabludoff 1998).  Others
(Hickson, Kindl, \& Auman 1989; Williams \& Rood 1987) also have found
that approximately 50\% of galaxies in compact groups are
bulge-dominated systems.  These studies show that the relationship
between environment and galaxy type is still unclear.  A limitation to
understanding the link between galaxy morphology and environment in
all these studies is how the galaxies are typed.  Typing was usually
done by eye and information like star-forming regions, tidal tails,
and shells were noted but not included in a quantitative analysis.  In
addition, not only are there variations dependent on the author
({\it e.g.} Fabricant, Franx, \& van Dokkum 2000) but visual typing is
difficult to do consistently for a sample with hundreds of galaxies.
An important first step in improving our understanding of what
determines galaxy morphology in different environments is to type
galaxies in a uniform and reproducible way so that we can compare
these galaxies in a quantitative manner.

An important aspect of understanding how galaxy morphology depends on
environment is observing how the early-type fraction in groups varies,
if at all, with position {\it within} the group.  Although the
morphology-density relation has been well-established for clusters
(Dressler 1980; Whitmore \& Gilmore 1991) and has been studied for
groups found in redshift surveys (Ramella $et~al.$ 1999; Postman \&
Geller 1984), the former is defined for regions of much higher density
and larger physical volume than that of poor groups and the latter is
defined by the {\it average} galaxy density of a group rather than as
a function of position {\it within} the group.  The morphology-radius
($\sim$ density) relation has not been examined in poor groups,
particularly those with extended X-ray emission that are likely to be
real, bound systems.  Our large sample of galaxies projected within
the virial radii of the poor groups, coupled with the quantitatively
derived morphologies, makes it possible to test if a morphology-radius
($\sim$ density) relation also exists in these groups.

We extend our analysis by testing how environment affects galaxy
formation and/or evolution by observing whether the star formation
levels and degree of morphological disruption in these galaxies
change with environment.  We address these questions in a detailed
manner by comparing only galaxies of the same morphological type.  In
past studies, it was not possible to make this clean a comparison
since we lacked a uniform classification scheme and a sample large
enough to be cut into statistically meaningful subsamples.  With our
large sample of spectroscopically-confirmed group members and
quantitative measurement of physical galaxy characteristics, however,
we now can use the bulge fraction to roughly isolate group galaxies of
a particular, narrow class and to then compare their current star
formation and asymmetry to their counterparts in the field.

In this paper, we characterize 171 galaxies in six nearby, X-ray
detected, poor groups and 18 galaxies in the field using a new
automated program that fits a PSF-convolved 2D surface brightness
profile to an image of each galaxy and searches $\chi^2$ space for the
best model (GIM2D; Simard $et~al.$ 2000; Simard $et~al.$ 1999; Marleau
\& Simard 1998).  For each group, we have 15 to 38
spectroscopically confirmed members ranging in luminosity from dwarfs
to giants ($-13.7\geq M_R-5$log$h\geq -21.9$) and in Hubble type from
irregulars and Sd's to S0's and ellipticals.  By fitting two component
profiles, we measure galaxy properties such as bulge fraction,
half-light radius, and the fraction of light in an asymmetric
component.  These structural parameters allow us to type the galaxy as
bulge or disk-dominated and to define its asymmetry in a quantitative
and reproducible manner.  Assuming a simple model, {\it e.g.} a de
Vaucouleurs (de Vaucouleurs 1948) bulge and exponential disk, we
determine the goodness of each fit and check the robustness of our
results for galaxies imaged at different effective resolutions.  We
then compare the distributions of structural parameters and their
relationship to current star formation (as measured by the 3727 \AA
~[OII] emission line) for the group and field samples.

The sections of this paper are organized in the following manner: The
data and the fitting technique are described in \S2.  Also included in
\S2 is an explanation of how we move all the galaxies to a common
redshift ($cz_0 = 9000$\kms) and effective resolution
(0.87$h^{-1}$ kpc) so that the samples may be directly compared.  In
\S3, we present the models and structural parameters fit by GIM2D for
the group and field galaxies at their original and common redshift.
We also test for relationships among the bulge fraction, asymmetry,
and the current star formation, discussing differences between the
group and field populations.  In \S4, we discuss possible explanations
for these differences.  Our conclusions are summarized in \S5.  In
this paper, we use H$_0=100 h$\kms~Mpc$^{-1}$ and q$_0=0.5$.

\section{Observations and Analysis}

\subsection{Observations}

We obtained images of the six galaxy groups (HCG 42, HCG 62, NGC 2563,
NGC 3557, NGC 4325, and NGC 5129) with the Tektronix $2048^2$ CCD and
Swope $40''$ telescope combination at Las Campanas Observatory during
October 1996 and February 1997.  Each image has a $\sim 23.8'$ field
of view, and we constructed a $3\times 3$ mosaic covering a
$1^{\circ}\times1^{\circ}$ field for each group.  The pixel scale is
$0.''696$.  Each of the nine tiles in the mosaic overlap by $\sim5'$
with an adjacent tile.  With the Kron-Cousins R filter, each tile was
exposed for a total time of 300 seconds.  The seeing varied for each
group, ranging from $1.''2$ to $2.''6$.  We reduced the images using
standard IRAF techniques.

We obtained complimentary spectra for group and field galaxies with
the multi-fiber spectrograph (Shectman $et~al.$ 1992) and the
2D-Frutti detector mounted on the du Pont 2.5 m telescope at the Las
Campanas Observatory.  The spectra in this paper are a subset of those
in Zabludoff \& Mulchaey (1998a; 2000) and are described there.  The
spectra have a pixel scale of $\sim 3$\AA, a resolution of $\sim
5-6$\AA, and a wavelength range of $3500-6500$\AA.  The [OII]
equivalent widths were measured in the same manner as in Zabludoff
$et~al.$ 1996.  At $m_R=16$, the galaxy groups are 80--100\%
spectroscopically complete (Zabludoff \& Mulchaey 2000).  As the
completeness function drops beyond $m_R\sim17$, we consider only
galaxies with $m_R<18$; the absolute magnitude range for the sample is
then $-13.7\ge M_R-5$log$h \ge -21.9$.  The absolute magnitude ranges
are consistent for the group and field samples (Fig.~\ref{fig3}, top
panel).  Nevertheless, in comparing the groups with the field, we test
our results for both the full absolute magnitude range and
$M_R\leq-17+5$log$h$ subsamples to account for any dependence on galaxy
luminosity.

The group properties are listed in Table 1.  For the field sample,
there are 18 galaxies with $m_R<18$ and $cz_0<9000$\kms; the field
galaxies were serendipitous observations in the fore/background of the
galaxy groups.  The limit of $cz_0<9000$\kms~is the redshift to which
we move our entire sample (group and field galaxies) for comparison
(see \S2.3).  

\placetable{table1}

\subsection{Structural Measurements}

We use the program GIM2D (Marleau \& Simard 1998; Simard $et~al.$
1999; Simard $et~al.$ 2000) to find the best-fit PSF-convolved, 2D
bulge+disk models to the surface brightness profiles of 171 group and
18 field galaxies.  The program has a maximum of 12 fitting
parameters: the flux ($F_{total}$) in the model integrated to
$r=\infty$; the bulge fraction $B/T \equiv F_{bulge}/F_{total}$; the
semi-major axis effective radius of the bulge $r_e$; the bulge
ellipticity $e\equiv(1-b)/a$ where $a$ and $b$ are the bulge
semi-major and semi-minor axes respectively; the bulge position angle
$\phi_b$; the semi-major axis exponential disk scale length $r_d$; the
inclination of the disk $i$ ($i \equiv 0$ for face-on); the disk
position angle $\phi_d$; the subpixel $dx$ and $dy$ offsets of the
galaxy's center; the residual background level $db$; and the
S\'{e}rsic index $n$ of the bulge.  Both $\phi_b$ and $\phi_d$ are
measured clockwise from the positive $y$-axis of the image.  Twelve
parameters is a maximum as one or more parameters can be frozen to
their initial values if necessary depending on the scientific goals
being pursued.  The best-fit parameters and their confidence intervals
are determined using the Metropolis algorithm (Metropolis $et~al.$
1953; Saha \& Williams 1994) which uses the $\chi^2_{\nu}$ test to
determine the region of maximum likelihood in the multi-parameter
space.

The bulge profile is defined as
\begin{equation}
\Sigma(r) = \Sigma_e~exp\left\{ -k[(r/r_e)^{1/n} - 1] \right\}
\label{bulge-prof}
\end{equation}
\noindent where $\Sigma(r)$ is the surface brightness at $r$ along the 
semi-major axis, and $\Sigma_e$ is the effective surface brightness.  
This bulge profile is also known as the S\'{e}rsic profile (S\'{e}rsic 
1968).  The parameter $k$ is equal to $(1.9992n - 0.3271)$, a value 
that defines $r_e$ to be the projected radius enclosing half of the 
light in the bulge component (Capaccioli 1989).  The classical de 
Vaucouleurs profile is a special case of Equation~\ref{bulge-prof} with 
$n = 4$.  

The disk profile is defined as
\begin{equation}
\Sigma(r) = \Sigma_0~exp(r/r_d)
\label{disk-prof}
\end{equation}
\noindent where $\Sigma_0$ is the central surface brightness.  We note 
as Simard $et~al.$ (1999, 2000) do that the bulge/disk nomenclature
adopted here to describe our surface brightness models may not reflect
the internal kinematics of its components.  A ``bulge'' may not be a
centralized, dynamically hot spheroid but a central starburst.
Similarly, a ``disk'' may not necessarily be a cold, co-rotating
population.  Some systems such as dwarf ellipticals are made up of a
dynamically hot component with an exponential surface brightness
profile.

Recent work (Courteau, de Jong, \& Broeils 1996; Andredakis 1998)
advocates a combination of two $n = 1$ exponential components to fit
the surface brightness profile of late-type spirals.  It is unclear
from our analysis, however, that this profile is more appropriate than
the classical de Vaucouleurs bulge plus exponential disk for the
galaxies in our sample.  We do fit a double exponential profile to
galaxies in all six group mosaics and compare the results to those of
the de Vaucouleurs bulge plus exponential disk models.  Both profiles
fit our galaxies equally well.  Only 6/189 galaxies are marginally
better fit by a double exponential profile; these range in magnitude
from $15<m_R<17.4$ and in bulge fraction from
$0\leq(B/T)_{deV}\leq0.9$.  In the absence of a clear preference
between the two bulge profiles and knowing that bright ellipticals and
the bulges of early-type spirals are well-fit by a de Vaucouleurs
profile (Andredakis 1998; Andredakis, Peletier, \& Balcells 1995), we
adopt this bulge profile plus an exponential disk as our canonical
fitting model for the sake of continuity across the full range of
morphological types.

Before GIM2D can fit a 2D model to the surface brightness of a given
galaxy, the isophotal area of the galaxy and the point spread function
(PSF) of the image at the location of the galaxy must first be
determined.  To define the isophotal area, we use the galaxy
photometry package SExtractor V2.0 (Bertin \& Arnouts 1996) with a
detection threshold of $\mu=24.05$ mags/$\Box''$, a minimum deblending
contrast parameter of 0.005, and two background mesh sizes
corresponding to galaxies brighter and fainter than $m_R = 13$
(200 and 64, respectively).  A variable mesh size is necessary since
SExtractor can mistake the flux of extended objects for sky if the
mesh size is too small and can thus over-subtract the sky from the
object.  We use the stand-alone version of DAOPHOT (Stetson 1987) to
measure and generate a PSF for each galaxy.  The PSF varies across the
image, and these PSF variations are mapped as a function of position
using stars located throughout the images.  PSF models are then
generated with DAOPHOT at regular intervals across the chip (every 50
pixels), and the PSF model closest to each galaxy is selected for its
GIM2D analysis.

By fitting models to the surface brightness distribution of these 
galaxies, we measure the structural properties $B/T$, $r_e$, $r_d$, 
$\phi_b$, $\phi_d$, $i$, and the half-light radius $r_{1/2}$.  The 
semi-major axis half-light radius is computed by integrating the sum 
of Equations~\ref{bulge-prof} and~\ref{disk-prof} to $r = \infty$.  
The asymmetric image residual flux is quantified by GIM2D using the 
$R_{A}$ index (Schade $et~al.$ 1995) defined as
\begin{eqnarray}
R_A & = & (R_A)_{raw} - (R_A)_{bkg} \nonumber \\
& = & {{\displaystyle\sum_{i,j} {1\over{2}}| R_{ij} - 
R_{ij}^{180}|} \over{\displaystyle\sum_{i,j} I_{ij}}} - {{\displaystyle\sum_{i,j} 
{1\over{2}}| B_{ij} - 
B_{ij}^{180}|} \over{\displaystyle\sum_{i,j} I_{ij}}}
\label{ra-index}
\end{eqnarray}	
where $R_{ij}$ is the flux at $(i,j)$ in the residual image, 
$R_{ij}^{180}$ is the flux in the residual image rotated by 
$180^{\circ}$, and $I_{ij}$ is the flux in the original image.  
Following Marleau \& Simard (1998), $R_A$ is measured within $r = 2 
r_{1/2}$.  The second term $(R_A)_{bkg}$ in Equation~\ref{ra-index} is 
a statistical correction for background noise fluctuations.  Since 
$(R_A)_{raw}$ involves taking absolute values of pixel fluxes, it will 
yield a positive signal even in the sole presence of noise.  The 
background correction, $(R_A)_{bkg}$, is computed over pixels flagged 
as background pixels in the SExtractor segmentation image.  The 
$B_{ij}$'s are background pixel values in the residual image, and the 
$B_{ij}^{180}$'s are background pixel values in the residual image 
rotated by $180^{\circ}$.  The background correction is computed over 
a background pixel area equal to the pixel area over which 
$(R_A)_{raw}$ is computed.  Given the statistical nature of 
$(R_A)_{bkg}$, there will be cases when a galaxy is faint enough 
compared to its background noise that $R_A$ may take on small negative 
values in exactly the way as the difference of two values of 
$(R_A)_{bkg}$ computed from different regions of the sky may be 
negative.

In addition to $R_{A}$, we measure the total residual fraction of
light $R_T$ by taking the pixels assigned to the galaxy by SExtractor
and creating a mask that is applied to the original and model images.
For a positive-definite residual fraction, the model is subtracted
from the original and the absolute value of the differences are
summed.  To account for the sky, the same number of sky pixels as
galaxy pixels are summed in the same manner and subtracted from the
total galaxy residual.  We use the total residual fraction of light as
a gauge of the model's goodness of fit but note that, for fainter
galaxies, the error in $R_T$ is dominated by the error in the flux.  

\subsection{Moving Galaxies to a Common Redshift}

In our analysis, individual groups are compared with each other and
with the field to test for differences among the group and field
populations.  To ensure that systematic biases in our quantitative
classification affect all galaxies in our sample in the same way, we
must observe all galaxies at the same effective resolution.  As the
galaxies lie at different distances and are imaged at different
seeings, we artificially adjust each galaxy image to a common redshift
and seeing by first reducing the galaxy's flux, changing the seeing as
the atmosphere then does, and finally binning the flux as a CCD
detector does.  This sequence thus mimics the real observations.  The
effective resolution is the physical radius of the seeing disk and is
determined by a combination of the distance (redshift) and seeing.
For example, the seeing in an image of the most distant group in our
sample, NGC 4325 ($cz=7558 \pm 70$\kms; Zabludoff \& Mulchaey 1998a)
is $1.''3$ and has an effective resolution of 0.46\hi kpc.  By moving
the galaxies to a common, higher redshift, we also test the robustness
of our results by comparing the galaxy models before and after the
galaxies are moved back and the physical resolution of the image
degraded.

We select a common redshift of $cz=9000$\kms, a slightly higher
redshift than our most distant group. We first fade the galaxies in
each group by
\begin{equation}
F=\left[ \frac{D_L(z_{9000})}{D_L(z_{fig1})} \right]^2
\end{equation}
where $F$ is the fading factor, and $D_L(z_{9000})$ and
$D_L(z_{fig1})$ are the cosmological luminosity distances to
$cz=9000$\kms~and the group's redshift respectively.  Since the
seeing for each image is different, we convolve each image with a
gaussian whose $\sigma$ is
\begin{equation}
\sigma = \frac{[ (FWHM_T)^2 - (FWHM_I)^2]^{1/2}}{2.35}
\end{equation}
where FWHM$_I$ is the image's original full-width at half-maximum and
FWHM$_T$ is the target full-width at half-maximum.  FWHM$_T$ is set by
NGC 5129, the group with the least effective resolution (0.87\hi kpc);
FWHM$_T=2.''1$ for $cz=9000$\kms.  The images then are binned
(conserving flux) by
\begin{equation}
B=\left[ \frac{D_A(z_{9000})}{D_A(z_{fig1})} \right]
\end{equation}
where $B$ is the size of the bin in pixels, and $D_A(z_{9000})$ and
$D_A(z_{fig1})$ are the cosmological angular distances to $cz=9000$
\kms~and the group's redshift respectively.  By fading and binning the
images, the sky noise is decreased; we add $\sim10$ counts of Poisson
noise to the refluxed and binned images so that the sky noise is the
same as in the original images.  The same process is applied to each
field galaxy.

\section{Results} 

\subsection{The Models}

We fit a de Vaucouleurs bulge with exponential disk model to the group
and field galaxies at their original redshifts and at a common
redshift of $cz=9000$\kms.  Table 2 lists the reference number,
redshift, absolute magnitude ($M_R$), [OII] equivalent width (EW),
bulge fraction ($B/T$; average error $\sim0.1$), half-light radius
($r_{1/2}$\hi kpc; average error $\sim10$\%), asymmetry parameter
($R_A$; average error $\sim0.02$), and fraction of residual light
($R_{T}$; average error $\sim0.02$) for the field galaxies at their
observed ($cz_0$) and common ($cz=9000$\kms) redshifts; Table 3 lists
the same parameters for the group galaxies.  In Fig. \ref{fig1}a, we
show all the galaxies (189) in the $1^{\circ}\times1^{\circ}$ mosaic
for the six galaxy groups.  The pixel area in each galaxy thumbnail is
12 times the galaxy's isophotal area as defined by SExtractor.  In
these images, the seeing ranges from $1.''3$ to $2.''6$.  The galaxies
range in magnitude from $-13.7+5$ to $-21.9+5$log$h$.  $M_R$ is listed
in the upper left corner of each thumbnail and the reference number in
the lower left.  Fig. \ref{fig1}b contains the best-fit de
Vaucouleurs with exponential disk model for each galaxy; $B/T$ is
listed in the upper left corner of each thumbnail.  The galaxies span
the range in bulge fraction ($0\leq B/T\leq 1$).  Fig. \ref{fig1}c
contains the residual images created by subtracting the best-fit
models from the original images.  Included in these thumbnails are the
asymmetry parameter $R_A$ and total residual fraction $R_T$.

In Fig. \ref{fig2}a, the same galaxies as in Fig. \ref{fig1}a
are shown but after being moved to $cz=9000$\kms~and the same
effective resolution (0.87\hi kpc).  As in Fig. \ref{fig1}, the
thumbnails include $M_R$, $B/T$, $R_A$, $R_T$, and reference number.
The effective resolution of 0.87\hi kpc at $cz=9000$\kms~is set by
NGC 5129, the group with the least effective resolution given its
combination of observed seeing ($2.''6$) and distance ($cz=6998\pm51$
\kms).  As expected, fading the galaxies, changing the PSF slightly,
and rebinning the image has diminished many of the features seen
prominently at $cz_0$.  In most cases, especially with small galaxies,
$R_T$ has decreased due to the loss of spatial resolution.  In a few
extreme cases, $B/T$ can change by as much as 0.59 (HCG 42-106),
$R_{T}$ as much as 0.30 (HCG 62-18), and $R_A$ as much as 0.16 (HCG
62-18).  We examine such changes more quantitatively in the next
section.

\placetable{table2}
\placetable {table3}

\placefigure{fig1}

\placefigure{fig2}

\subsubsection{How Good Are the Models?}

From the 2D models fit by GIM2D to the surface brightness
distributions of the galaxies in our sample, we see that most galaxies
are fit well with the combination of a de Vaucouleurs bulge and
exponential disk profile (cf. Figures \ref{fig1}b and
\ref{fig2}b).  Of the 189 galaxies in our sample, less than 10\%
of the galaxy's light remains in the residual fraction for 158
galaxies (84\%) at $cz_0$ and 171 galaxies (91\%) at $cz=9000$\kms.
Galaxies with high residuals (Figures \ref{fig1}c and
\ref{fig2}c) usually have bars, spiral arms, concentrated
star-forming regions, warped disks, and tidal tails (see NGC 2563-18,
NGC 5129-95, NGC 3557-5, NGC 2563-23, and NGC 2563-9 respectively),
features that are impossible to fit with any smooth profile.  We
measure morphological asymmetries with $R_A$, a quantitative measure
that facilitates comparison among the galaxies and their spectra.

A technical detail is how GIM2D tends to overestimate the disk
component if there are bright features associated with it, {\it e.g.} spiral
arms and HII regions (see NGC 2563-19, NGC 3557-5, and NGC 5129-6). This is
a problem for all two-dimensional fitting programs using unsymmetrized
images and affects mainly low $B/T$ galaxies.  Bright features in the
disk cause GIM2D to overestimate the contribution from the disk
component and increase its scale length $r_d$.  Thus, the galaxy's
disk is oversubtracted and its $B/T$ underestimated.  These cases are
easily identified by their high residual and asymmetry fractions.

Using the residual fraction of light ($R_{T}$) as a guide to the
goodness of the models, we plot in Fig. \ref{fig3} $R_{T}$ as a
function of absolute magnitude for both the original and redshifted
galaxies.  The dotted line denotes $R_{T}=0$.  The bottom panel shows
the difference $[(R_T)_{cz_{0}}-(R_T)_{cz_{9000}}]$ as a function of
$M_R$ for the group and field galaxies.  The group and field galaxies
are fit equally well with a de Vaucouleurs with exponential disk model
across the range in absolute magnitude.  As galaxies are redshifted
and the seeing worsened, we would expect the fits to somewhat
``improve'' ({\it i.e.}, have smaller $R_T$) on average due to the
loss of effective resolution.  A comparison of the $cz_0$ and $cz =
9000$\kms~results shows that the fits to intrinsically faint (in
general, small) galaxies do improve slightly when the galaxies are
redshifted; the average $[(R_T)_{cz_{0}}-(R_T)_{cz_{9000}}]$ is $0.02
\pm 0.01$ for galaxies fainter than $M_R = -20 + 5$log$h$.  For
brighter (in general, bigger) galaxies, the ratio of the seeing disk
to the galaxy size is relatively unaffected, and the number of pixels
per galaxy remains large, from $cz_0$ to $cz = 9000$\kms.

\placefigure{fig3}

\subsection{Morphological Classifications}

\subsubsection{At the Observed Redshift ($cz_0$)}

In Fig. \ref{fig4}, we compare the measured bulge fractions ($B/T$)
for the field and group galaxies to their published Hubble types (from
the NASA Extragalactic Database) to check for correspondence.  The
visually classified galaxies are split into four bins: Ellipticals
(E); S0's; Spirals/Barred Spirals (S/SB); and Irregular/Peculiar
(Irr/Pec).  Galaxies with significant asymmetry ($R_A\geq0.05$) are
circled.  We note that only 24\% of the entire sample has been
visually typed and that the typing was done by different authors.  It
is clear from Fig.~\ref{fig4} that we cannot match the four
morphological bins with four $B/T$ bins.  Instead, we use $B/T$ to
split the galaxies into ``early'' and ``late'' types, where the former
are bulge-dominated galaxies and the latter are disk-dominated
galaxies.

\placefigure{fig4}

Comparing the Hubble type to the measured $B/T$ for the non-redshifted
and redshifted samples, we define the break between early and
late-types at $B/T=0.4$.  As seen in Fig.~\ref{fig4}, the
bulge-dominated galaxies include a range of Hubble types, whereas the
disk-dominated systems correspond to Hubble type spirals or
irregulars.  One discrepant galaxy is a field galaxy
($M_R=-16.4+5$log$h$) classified in the literature as a dwarf
elliptical but whose best-fit model is a pure disk ($B/T = 0$).  Dwarf
ellipticals have been found, however, to have exponential profiles
(Ferguson \& Bingelli 1994).  Interestingly, the same galaxy was
classified as a spiral by Zabludoff \& Mulchaey (1998a; 1999).  The 12
galaxies with S/SB Hubble types but with $B/T>0.4$ were manually
checked by two of the authors (K. T. and L. S.): ten have no visible
disk and a de Vaucouleurs type profile, one has a prominent thick bar,
and the last has a faint disk with a bright bulge.  While the measured
bulge fraction does not correspond to the published Hubble type for
these 12 galaxies, the GIM2D models do match the galaxies' images
well.

\subsubsection{At a Common Redshift ($cz=9000$\kms)}
 
The lower panel of Fig.~\ref{fig4} shows $B/T$ and the Hubble types
for the redshifted sample.  The distribution is similar to that at
$cz_0$.  We test the robustness of our results by examining the change
in bulge and asymmetry fraction for galaxies in the two morphological
bins.  Fig.~\ref{fig5} shows how $B/T$ and $R_A$ change when
group galaxies are redshifted to $cz=9000$\kms.  The galaxies are
divided into bulge ($B/T\geq0.4$) and disk-dominated ($B/T<0.4$)
galaxies.  They are binned in $cz$ with the median in each bin and the
corresponding asymmetric $1\sigma$ confidence limits shown.  There is
no significant change in the median $B/T$ for either bulge or
disk-dominated group galaxies, even though some galaxies have been
moved up to four times further away and the effective resolution
degraded.  $B/T$ is stable for the bulk of the sample but a few
outliers change significantly by either decreasing or increasing in
$B/T$: 13/87 of $(B/T)_{cz_0}<0.4$ galaxies have
$(B/T)_{cz_{9000}}\geq0.4$ while 11/85 of $(B/T)_{cz_0}\geq0.4$ have
$(B/T)_{cz_{9000}}<0.4$.  Classifying galaxies by their bulge fraction
is thus robust on average for the sample but $B/T$ ({\it i.e.},
classification) for individual galaxies may change by up to 0.6.

\placefigure{fig5}

\subsection{The Fraction of Bulge-Dominated Galaxies ($B/T\geq0.4$)}

In Fig.~\ref{fig6} we show the bulge fraction distributions for the
group and field galaxies at their original and common redshift.  The
$B/T$ distribution in the groups is different than that in the field.
Application of the Kolmogorov-Smirnov test (Press $et~al.$ 1992) rules
out a common parent distribution with $>95$\% certainty at $cz_0$ and
$cz=9000$\kms.  As we do not sample the same physical volume in all
the groups, however, we test this result by including only galaxies
within $R=0.25$\hi Mpc of their group centers.  This projected
distance limit is selected based on the radius corresponding to the
$1^{\circ}\times1^{\circ}$ field for the closest group, NGC 3557.
Again, a common parent distribution for the field and groups is ruled
out with $>95$\% certainty.

\placefigure{fig6}

With the separation between early and late-type galaxies defined at
$B/T=0.4$, we find that the groups have a higher fraction of
early-type galaxies than the field (86/171=50\% vs. 3/18=17\% at $cz =
9000$ km s$^{-1}$).  This result is roughly consistent with those
obtained by visual classification of compact group galaxies (cf. Rood
\& Williams 1989; Hickson, Kindl, \& Huchra 1988; Zabludoff \&
Mulchaey 1998).  A high early-type fraction is found in the individual
groups as well as for the entire group sample.  Moving the galaxies
from $cz_0$ to a common redshift and effective resolution does not
change the result significantly even though the effective resolution
is decreased by as much as five for some galaxies.  Including only
galaxies within 0.25\hi Mpc of the groups' centers only increases the
early-type fraction.  These values are listed in Table 4.

\placetable{table4}

We also test that the early-type fraction in groups and in the field
are significantly different in the following manner.  We calculate the
probability that the group and field early-type fractions are the same
by using the cumulative binomial probability distribution to determine
whether there is a ``parent'' early-type fraction consistent with both
the field and group early-type fractions.  For example, suppose that
$k_1/n_1$ is the observed fraction of group galaxies that are
early-types and $k_2/n_2 < k_1/n_1$ is the observed fraction of field
galaxies that are early-types.  For each parent probability $p = 0$ to
1, we calculate the probability $P_1$ that {\it at least} $k_1$ of
$n_1$ group galaxies are early-types,
\begin{equation}
P_1 \equiv \sum_{j=k_1}^{n_1} (^{n_1}_j )p^j(1-p)^{n_1-j} = 
I_p(k_1,n_1-k_1+1), 
\end{equation}
where $I_p$ is the incomplete beta function (Press $et~al.$ 1992).
For the same $p$, where $p$ is the probability of event occurring per
trial, we calculate the probability $P_2$ that {\it at most} $k_2$ of
$n_2$ field galaxies are early-types,
\begin{equation}
P_2 \equiv 1 - I_p(k_2+1,n_2-k_2).  
\end{equation}
The product of the probability distributions $P_1(p)P_2(p)$ is the
joint probability distribution whose maximum tells us the likelihood
that both $k_1/n_1$ and $k_2/n_2$ were drawn from the same parent
fraction.  The probability that the early-type fraction of the group
galaxies is the same as that of the field galaxies is 0.0019 ({\it i.e.},
significantly different at the $> 95$\% level).

A possible caveat is the incompleteness of our sample at fainter
magnitudes ($M_R>-17$).  To test the result's robustness, we impose a
magnitude limit of $M_R\leq-17$ on the group and field galaxies.  This
decreases the group sample to 132 galaxies and the field to 14.  With
this limit, the galaxy groups are 80-100\% complete.  Again, we find
that groups have a significantly higher fraction of early-type
galaxies than the field: 67/132 (51\%) to 3/14 (22\%) at $cz=9000$
\kms.

\subsection{The Spatial Distribution of Bulge-Dominated Galaxies}

The $B/T$ parameter permits us to test whether a morphology-radius
($\sim$ galaxy density) relationship, similar to that of galaxy
clusters (Dressler 1980; Whitmore \& Gilmore 1991), exists in poor
groups.  In the five groups for which our sampling extends beyond
projected radii of $R>0.25$\hi Mpc (HCG 42, HCG 62, NGC 2563, NGC
4325, and NGC 5129), the early-type fraction decreases when all the
group galaxies in the $1^{\circ}\times1^{\circ}$ field are included.
In Fig.~\ref{fig7}, the bulge fraction measured at $cz=9000$\kms~
for all group members (171) is shown as a function of their projected
radius from the fiber field center ($0\geq R \geq0.6$\hi Mpc).  The
least-squares fit to the data and the average $B/T$ for each 50\hi kpc
bin both show a decrease in $B/T$ with increasing projected distance
from the group center.  Using the Spearman rank test (Press $et~al.$
1992), we find the anti-correlation between $R$ and $B/T$ to be
$>95$\% significant.

\placefigure{fig7}

The morphology-radius relation, high fraction of early-types, and
diffuse X-ray halos of these groups also are observed in galaxy
clusters.  These points suggest that X-ray detected poor groups may
form as low mass analogs of rich clusters (cf. Mulchaey \& Zabludoff
1998).

\subsection{The Fraction of Asymmetric Galaxies ($R_A\geq0.05$)}

As discussed in \S2.2, GIM2D also quantifies galaxy asymmetry ($R_A$).
Asymmetric features including tidal tails, shells, warped disks, and
asymmetrically distributed HII regions increase $R_A$.  After visually
inspecting the galaxies and their residual images, we find that all of
the galaxies with $R_A\ge0.05$ have a visually obvious asymmetric
residual light distribution (see \ref{fig1}c and \ref{fig2}c).
Therefore, we adopt $R_A\ge0.05$ as the lower limit for galaxies that
we classify as having high galaxy asymmetry.

At $cz=9000$\kms, the fraction of asymmetric galaxies is 19/171
(11\%).  We test for any radial or luminosity dependence in the
asymmetry fraction by imposing the same radial and magnitude limits as
in \S3.3.  Including only galaxies within 0.25\hi Mpc of the group's
center does not change significantly the asymmetry fraction in the
groups, nor does applying a magnitude limit of $M_R\leq-17$.  It is
not feasible to compare the asymmetry fraction of individual groups
because we are limited by the small number of asymmetric members in
each group.

Comparing the results for the original and redshifted galaxies, we
observe a slight decrease in the ability to recover asymmetric
features with decreased effective resolution (Fig.~\ref{fig5}):
the mean change is $0.02\pm0.01$ at all $cz_0$ for the entire group
sample.  We note that 13/24 (54\%) of the galaxies with
$(R_A)_{cz_0}\geq0.05$ are not considered significantly asymmetric
after they are moved to $cz=9000$\kms.  This result is due to the
decrease in effective resolution and the loss of morphological
information.  Of the 147 $cz_0$ galaxies with $R_A<0.05$, however,
eight (5\%) are considered significantly asymmetric at $cz=9000$\kms.
In this case, the fading of the galaxy images makes asymmetries more
prominent by reducing the contribution from the lower surface
brightness, symmetric component.  As in the case of $B/T$, there are
as many outliers whose $R_A$ increases as there are whose $R_A$
decreases such that the average change in $R_A$ is fairly small, even
though $R_A$ for individual galaxies can change by as much as 0.16.
The asymmetry index measured does correspond to the galaxy's image but
the question of whether the image truly reveals the galaxy's
properties becomes increasingly important with increasing redshift.

In Fig. \ref{fig8}, the distribution of $R_A$ with respect to $B/T$
for the groups and the field at $cz=9000$\kms~is shown.  The lower
panel shows the fraction of group galaxies with high galaxy asymmetry
($R_A\geq0.05$) for each $B/T$ bin.  There is no correlation between
$R_A$ with $B/T$ in these groups (the results are similar at $cz_0$).
Group members with high galaxy asymmetry are not even predominantly
disk-dominated systems: 10/19 of the significantly asymmetric group
galaxies at $cz=9000$\kms~have $B/T\geq0.4$.

Of the asymmetric disk-dominated systems, 8/9 (89\%) are star forming
(see \S3.6). Thus, their high $R_A$ values most likely result from (1)
asymmetrically distributed HII regions, (2) interaction-induced
features, and/or (3) the presence of a close companion inside the
$2r_{1/2}$ aperture used to calculate $R_A$ (no close companions are
visible, however).  In contrast, none of the asymmetric
bulge-dominated galaxies at $cz=9000$\kms~are star-forming.  Thus,
their high $R_A$ value may result only from the latter two
explanations.  Unlike the disk-dominated systems, 3/10 asymmetric
bulge-dominated galaxies do have close companions (HCG 62-1, HCG 62-4,
and HCG 62-18).

\placefigure{fig8}

To test for differences between the asymmetry fraction of the groups
and the field, we consider only a narrow class of disk-dominated
galaxies ($B/T<0.2$).  Because group galaxies have a higher $B/T$ on
average than field galaxies (\S 3.3), selecting a narrow range of
$B/T$ allows us to first remove any dependence of $R_A$ on $B/T$
(however small).  The range $B/T < 0.2$ is chosen because it contains
most of the field sample and thus the statistics for a field/group
comparison are best within this $B/T$ range.  Using the cumulative
binomial probability distribution (\S 3.3), we find that for these
strongly disk-dominated galaxies the field asymmetry fraction
($6/13=46$\%) differs from that of groups ($7/52=13$\%) at the $>95$\%
confidence level (see Fig. \ref{fig9}).  Thus, not only do group
galaxies tend to have higher bulge fractions than field galaxies but
the most disk-dominated group galaxies are less asymmetric on average
than their counterparts in the field.

\placefigure{fig9}

\subsection{Current Star Formation}

\subsubsection{$B/T$ vs. [OII] Emission}

In the following discussion, we compare the [OII] emission with the
structural parameters measured using GIM2D for the composite group and
the field galaxies.  From the spectra, we have the [OII] flux estimate
and equivalent width (EW); we use the former to define a star-forming
galaxy and the latter as a measure of the star formation rate
(Kennicutt 1992; Balogh $et~al.$ 1999).  Note that we assume [OII]
emission indicates star formation and not AGN activity.  For
uniformity, we consider only results for the redshifted
($cz=9000$\kms) sample because we focus on the comparison of the
composite group with the field.  Figure~\ref{fig10} shows $B/T$,
[OII] EW, and the asymmetry $R_A$ versus absolute magnitude
($M_R$). From this figure, it is clear that $B/T$, [OII] EW, and $R_A$
vary with galaxy luminosity.  Therefore, it is possible to obtain
misleading results by comparing samples with different luminosity
ranges or by considering only the brightest galaxies.  We stress that
our group and field samples have similar absolute magnitude ranges,
and that including only galaxies brighter than $M_R=-17+5$log$h$ does
not affect our overall results.

\placefigure{fig10}

In Fig.~\ref{fig11} the [OII] EW is plotted against $B/T$ for the
group and field galaxies.  The upper panel shows the [OII] EW
distribution versus $B/T$.  The lower panel shows the fraction of
group galaxies with [OII] {\it flux} greater than $2\sigma$ where
$\sigma$ is the error in the flux.  Current star formation often is
associated with late-type galaxies (Hashimoto $et~al.$ 1998; Kennicutt
1992), and comparison of the [OII] emission to $B/T$ in our redshifted
sample agrees with this finding.  Of the 30 star-forming galaxies
([OII] flux $\geq2\sigma$), 22 are disk-dominated ($B/T<0.4$) systems
ranging in magnitude from $-16.3$ to $-20.3+5$log$h$.  Particularly
interesting, however, is the fraction of star-forming galaxies that
are bulge-dominated: 8 of 30 (27\%) star-forming galaxies ranging in
magnitude from $-15.5$ to $-20.9+5$log$h$ have $B/T\geq0.4$.  The
fraction of all bulge-dominated galaxies that are forming stars is
small (8/86; 9\%) but covers a large range in luminosity.

\placefigure{fig11}

These 30 star-forming galaxies are not concentrated in the group
centers but are 90-660\hi kpc away in projected radius.  In
comparison, quiescent group galaxies are 10-630\hi kpc from their
group center.  Application of the Kolmolgorov-Smirnov test on the two
radial distributions excludes a common distribution with $>95$\%
confidence.  Like Zabludoff \& Mulchaey (1998), we find that galaxies
with significant [OII] emission tend to lie further from the projected
group center than those with no significant emission.  We note,
however, that this result may be strongly correlated with the
morphology-radius ($\sim$ density) relation observed in these groups.
For example, a K-S test of the radial distributions for the most
disk-dominated ($B/T<0.2$) star-forming and quiescent group galaxies
(16 and 36 respectively) finds them to be indistinguishable.

\subsubsection{Asymmetry vs. [OII] Emission}

Disturbed galaxies often show signatures of current star formation in
their spectra (Young $et~al.$ 1996), suggesting that the mechanisms
that trigger bursts also can alter galaxy morphology.  These
starbursts are characterized by strong [OII] emission (Kennicutt
1992), which serves as a spectral snapshot of the galaxy's current
star-forming activity.  If asymmetries are related to star formation,
galaxies with strong [OII] emission should have higher $R_A$ values.
We test the link between recent star formation activity and the
morphological characteristics of our sample galaxies by comparing
their [OII] EW to the asymmetry parameter $R_A$.  In
Fig.~\ref{fig12} we show [OII] EW versus $R_A$ for bulge-dominated
($B/T\geq0.4$) and disk-dominated ($B/T<0.4$) galaxies in the groups
and the field at $cz=9000$\kms.

\placefigure{fig12}

There is a trend of increasing [OII] emission with increasing $R_A$
for {\it disk-dominated} galaxies; the trend is $>95$\% significant
using the Spearman rank test (Press $et~al.$ 1992).  Two possible
explanations of this correlation are: 1) an external cause such as a
physical connection between asymmetry and star formation so that if a
galaxy is morphologically disturbed by an external force like a merger
or tidal interaction, its star formation also is elevated or 2) an
internal cause such as asymmetrically distributed HII regions that
will increase $R_A$ (our code cannot distinguish between this case and
a tidally disrupted galaxy).

We do not find a significant correlation between [OII] emission and
high $R_A$ for {\it bulge-dominated} group galaxies.  We note,
however, that this does not mean there is no link between high $R_A$
and star formation for these galaxies.  A correlation may exist but
our ability to test for one is limited by the small fraction of these
galaxies with significant [OII] emission.  We also note that in
Fig.~\ref{fig12}, $B/T\geq0.4$ objects with large [OII] EW's and
low $R_A$ actually may be galaxies with central star-forming regions
rather than a true bulge.

To test for differences between the star forming fraction of the
groups and field, we again consider only objects in a narrow class of
late-types ($B/T<0.2$; see Fig. \ref{fig9}).  In doing so, we
maximize the statistics for a field/group comparison and remove any
dependence between [OII] emission and $B/T$.  In this $B/T$ bin, the
field and group star-forming galaxies have indistinguishable absolute
magnitude ranges with the K-S test.  Using the cumulative binomial
probability distribution (\S3.3), we find that for these strongly
disk-dominated galaxies, the fraction forming stars is significantly
higher ($>95$\% confidence) in the field than in the groups: 9/13
(69\%) compared to 16/52 (31\%).  In addition, the field's average
[OII] EW for these galaxies is twice that of the groups:
$14.3\pm0.5$\AA~compared to $7.4\pm0.4$\AA.  Not only do group
galaxies differ from the field in early-type fraction but for the most
disk-dominated galaxies, the galaxy asymmetry, star formation
fraction, and star formation rate are lower in groups
(Fig. \ref{fig9}).

\section{Discussion}

The group galaxy populations in our sample differ from the field
morphologically and in current star formation: 1) The early-type
fraction in groups is higher than the field. 2) For the most
disk-dominated systems, the fraction of galaxies that are asymmetric,
the fraction that are star-forming, and the average [OII] EW in the
galaxies are lower in groups than in the field.  The first point is
consistent with the predictions of standard biased galaxy formation
(White 1994), in which more massive (typically, bulge-dominated)
galaxies form in denser regions, and does not obviously require
subsequent environment-dependent galaxy evolution.  Possible
explanations for the second point include:

I.  Forming in regions of higher mass density than the field, groups
collapse earlier and their galaxies may form earlier as well.  Thus, a
group late-type may have consumed more of its gas by the current epoch
than a similar galaxy in the field, leading to less star formation and
more symmetry in the group galaxy.

II.  The environment of X-ray luminous groups might shut down star
formation (and make galaxies look more symmetric), although processes
such as ram pressure stripping are not thought to be as effective in
group environments as in clusters due to the lower average velocity
dispersions of groups (cf. Zabludoff \& Mulchaey 1998).  For some
rich clusters, such environment-dependent evolution mechanisms are
invoked to explain the lower average star formation activity in
cluster galaxies relative to the field (Poggianti $et~al.$ 1998;
Balogh $et~al.$ 1999).

III.  Interactions and mergers between galaxies, which are known to
enhance star formation (Young $et~al.$ 1996) and disrupt galaxy
morphology (Mihos 1995), may occur more frequently outside of X-ray
luminous groups and clusters at the current epoch.  Although these
groups are probable sites for interactions between galaxies because
their velocity dispersions are comparable to those of individual group
members (Barnes 1985; Aarseth \& Fall 1980), mergers that could boost
star formation and asymmetry are less likely today for two reasons.
First, the predominance of early-types in X-ray luminous groups
reduces the chance of any two gas-rich galaxies merging at the current
epoch.  Second, most of the group mass lies in a common halo rather
than with individual galaxies (Zabludoff \& Mulchaey 1998).  This mass
distribution reduces the cross-section of the galaxies and thus the
merger rate (Athanassoula, Makino, \& Bosma 1997).  In such a model,
if interactions occur they do so early in the group's evolutionary
history, and may be the reason for the predominance of quiescent,
undisturbed early-type galaxies in X-ray luminous groups today.

\section{Conclusions}

Using multi-object spectroscopy and wide-field CCD imaging of six
nearby X-ray detected, poor galaxy groups, we characterize the
morphology and current star formation of the group members and of a
comparison sample of field galaxies.  The large number of
spectroscopically confirmed members in each group allows us to sample
a wide range of galaxy types and luminosities, whereas most previous
group studies have been limited to only the few brightest members.  By
objectively classifying the galaxies into early and late-types using
their bulge fractions, quantitatively measuring the asymmetry of each
galaxy, and determining the [OII] emission line strength, we also
investigate the relationship between galaxy morphology and current
star formation for galaxies in these common, yet poorly understood,
environments.

We measure the structural parameters and morphologically classify 171
group members and 18 field galaxies. Surface brightness models are fit
to the galaxies at their observed ($cz_0$) and at a common ($cz=9000$
\kms) redshift.  By fitting 2D models to the galaxies at these two
redshifts, the composite of the group sample can be compared directly
to the field, and the robustness of the models is tested.

Our conclusions are as follows:

I.  A de Vaucouleurs bulge with exponential disk model fits well the
    surface brightness profile of the group and field galaxies in our
    sample.  Of the 189 galaxies in our sample, 172 (91\%) have
    residual fractions $<0.1$ once the model is subtracted (at
    $cz=9000$\kms).  Our ability to fit models to the data does not
    depend strongly on morphological type or luminosity.  The majority
    of galaxies with high residuals have bars, spiral arms, tidal
    tails, star-forming regions, and dust lanes, features that cannot
    be fit with any smooth profile.

II.  Morphologically classifying galaxies by bulge fraction ($B/T$) is
    fairly robust on average for the galaxies in our sample of nearby
    groups, even when the galaxy's redshift is increased by up to a
    factor of four and the effective resolution degraded by up to a
    factor of five.  The average asymmetry ($R_A$) of these galaxies
    decreases slightly at the higher redshift ($0.02\pm0.01$).  We
    note, however, that both $B/T$ and $R_A$ can change in either
    direction by as much as 59\% and 30\%, respectively, for individual
    galaxies.

III.  The fraction of early-type ($B/T\geq0.4$) galaxies in X-ray
    detected, poor groups is higher than the field ($\sim 50$\%
    vs. $\sim 20$\% at $cz=9000$\kms).  A common parent galaxy
    distribution for the group and field samples can be ruled out with
    $>95$\% certainty.

IV.  Groups have a morphology-radius ($\sim$ density) relation.  The
    average $B/T$ for the combined group sample decreases from the
    core out to $0.6$\hi Mpc.  The anti-correlation between between
    $B/T$ and radius is $>95$\% significant using the Spearman rank
    test.

V.  Current star formation, as characterized by [OII] emission, is
    correlated with high asymmetry ($R_A\geq0.05$) for disk-dominated
    ($B/T<0.4$) systems in the groups.  Two possible explanations for
    this correlation are: (1) A physical connection between morphology
    and star formation exists such that when group galaxies are
    disturbed by an external force, their star formation also is
    boosted.  (2) The fact that asymmetrically distributed HII regions
    will cause a high $R_A$ (and that our code cannot distinguish
    between this case and, for example, a tidally disrupted galaxy).
    While we cannot always identify the sources of high asymmetry in
    disky, star-forming galaxies, there is less ambiguity for
    quiescent early-types.  There is a non-negligible fraction of
    early-type group galaxies (10/86=12\%) that are both asymmetric
    and quiescent -- these are likely to be morphologically disturbed
    and/or to have a close companion.

VI.  For the most disk-dominated galaxies ($B/T<0.2$), the fraction
    with a high degree of asymmetry ($R_A\geq0.05$) is significantly
    higher in the field than in groups (6/13=46\% vs. 7/52=13\%,
    respectively), the fraction that are forming stars is
    significantly higher in the field (9/13=69\% vs. 16/52=31\%,
    respectively), and the average star formation rate is
    significantly higher in the field
    ($14.3\pm0.5$\AA~vs. $7.4\pm0.4$\AA, respectively).  These points
    suggest that although there are some highly disturbed-looking
    group galaxies in our sample, very late-type galaxies in
    X-ray detected groups are less star forming and asymmetric on
    average than similar galaxies in the field.

\acknowledgments

We are grateful to Katherine Wu for helpful discussions about
quantifying galaxy morphologies and Chuck Keeton for useful discussions
on statistics.  We thank the anonymous referee for his helpful
comments and quick response.  K. H. T. thanks Garth Illingworth for
his patience and Helmut Katzgraber for his unflagging support.  This
work was partially supported by NASA grant HF-01087.01-96A.  This
paper is based on observations made at Las Campanas Observatory,
Chile.

\newpage

\begin{deluxetable}{lrrrrrr}
\tablecolumns{7}
\tablewidth{0pc}
\tablecaption{Group Properties \tablenotemark{a}\label{table1}}
\tablehead{
\colhead{Group}    &    \colhead{$\alpha$ (2000.0)}   &
\colhead{$\delta$ (2000.0)} &       \colhead{N\tablenotemark{b}} &
\colhead{$\bar{v}$}   	&       \colhead{$\sigma$}  	&
\colhead{$D$\tablenotemark{c} (\hi kpc)}	}
\startdata
NGC2563 &  8 20 24.4 & +21 05 46 & 36 & 4775$\pm$65 & 336$\pm$44 & 811\\
NGC3557 & 11 09 58.5 & -37 23 03 & 15 & 2843$\pm$64 & 282$\pm$50 & 488\\
NGC4325 & 12 23 18.2 & +10 37 19 & 23 & 7558$\pm$70 & 265$\pm$50 & 1272\\
NGC5129 & 13 24 28.1 & +13 55 32 & 34 & 6998$\pm$51 & 304$\pm$43 & 1173\\
HCG 42 	& 10 00 13.1 & -19 38 24 & 25 & 3830$\pm$47 & 266$\pm$37 & 654\\
HCG 62 	& 12 53 12.4 & -09 12 12 & 38 & 4356$\pm$50 & 376$\pm$52 & 741\\
\tablenotetext{a}{The values for $\alpha, \delta, \bar{v}$, 
and $\sigma$ are from Zabludoff \& Mulchaey 1999.}
\tablenotetext{b}{Total number of group members in the $1^{\circ}
\times1^{\circ}$ field (Zabludoff \& Mulchaey 1999).}
\tablenotetext{c}{Physical diameter subtended by $1^{\circ}$ at the
group's mean recessional velocity (H$_0=100 h$ \kms Mpc$^{-1}$; q$_0=0.5$).}
\enddata
\end{deluxetable}

\begin{deluxetable}{lrrrrrrrrcrrrr}
\tabletypesize{\tiny}
\tablecolumns{14}
\tablewidth {0pc}
\tablecaption{Structural Parameters of Field Galaxies Measured at $cz_0$ 
and $cz=9000$ \kms \label{table2}}
\tablehead{
\multicolumn{5}{c}{}    & \multicolumn{4}{c}{$cz_0$}    &
\colhead{}          & \multicolumn{4}{c}{$cz=9000$ \kms} \\
\cline{6-9} \cline{11-14}   \\
\colhead{Field}     & \colhead{ID}  & \colhead{$cz_0$ (\kms)} &
\colhead{$M_R-5$log$h$\tablenotemark{a}}     & 
\colhead{[OII] EW (\AA)}      &
\colhead{B/T\tablenotemark{b}}
 &
\colhead{$r_{1/2}$\tablenotemark{c}~ (\hi kpc)}   & 
\colhead{$R_A$\tablenotemark{d}}   &
\colhead{$R_T$\tablenotemark{e}} &
\colhead{} &            \colhead{B/T} &
\colhead{$r_{1/2}$ (\hi kpc)}   & \colhead{$R_A$}   & \colhead{$R_T$} }
\startdata
H42 & 20 & 7611$\pm$37 & -19.8 & 6.9$\pm$0.7 & 0.44 & 2.94 & 0.08 & 0.19 &  & 0.01 & 1.08 & 0.02 & 0.07 \\
H42 & 25 & 8425$\pm$35 & -19.3 & 3.7$\pm$0.9 & 0.18 & 4.93 & 0.04 & 0.06 &  & 0.03 & 2.11 & 0.08 & 0.08 \\
H42 & 30 & 7555$\pm$12 & -18.4 & 34.8$\pm$2.0 & 0.39 & 2.98 & 0.06 & 0.12 &  & 0.19 & 1.38 & 0.02 & 0.02 \\
H42 & 131 & 7610$\pm$33 & -17.1 & 22.1$\pm$1.6 & 0.19 & 1.21 & 0.02 & 0.04 &  & 0.19 & 0.62 & 0.08 & 0.10 \\
H62 & 24 & 2266$\pm$40 & -18.4 & 13.7$\pm$1.5 & 0.15 & 5.34 & 0.08 & 0.19 &  & 0.13 & 9.74 & 0.05 & 0.15 \\
H62 & 42 & 7341$\pm$35 & -18.6 & 34.8$\pm$2.2 & 0.15 & 2.73 & 0.07 & 0.08 &  & 0.16 & 1.62 & 0.05 & 0.05 \\
N2563 & 15 & 6708$\pm$28 & -20.3 & 0.5$\pm$1.0 & 0.93 & 2.29 & 0.03 & 0.10 &  & 0.56 & 1.51 & 0.02 & 0.04 \\
N2563 & 17 & 6387$\pm$60 & -19.6 & 2.3$\pm$1.2 & 0.04 & 3.14 & 0.08 & 0.14 &  & 0.02 & 2.26 & 0.04 & 0.12 \\
N2563 & 22 & 6482$\pm$24 & -19.8 & 4.9$\pm$1.2 & 0.41 & 2.78 & 0.04 & 0.07 &  & 0.48 & 2.22 & 0.02 & 0.03 \\
N2563 & 48 & 6488$\pm$62 & -18.3 & 20.6$\pm$4.5 & 0.03 & 2.04 & 0.04 & 0.07 &  & 0.00 & 1.51 & 0.01 & 0.03 \\
N2563 & 86 & 6932$\pm$40 & -17.8 & 12.2$\pm$1.7 & 0.09 & 2.53 & 0.06 & 0.08 &  & 0.00 & 1.69 & 0.00 & 0.01 \\
N2563 & 88 & 2058$\pm$12 & -15.3 & 41.9$\pm$14.8 & 0.25 & 0.60 & 0.03 & 0.08 &  & 0.53 & 1.87 & 0.01 & 0.04 \\
N2563 & 152 & 2073$\pm$92 & -14.4 & -0.3$\pm$6.3 & 0.46 & 0.41 & 0.01 & 0.07 &  & 0.29 & 0.83 & 0.01 & 0.01 \\
N4325 & 23 & 2067$\pm$104 & -16.4 & -0.1$\pm$2.0 & 0.00 & 1.84 & 0.02 & 0.04 &  & 0.00 & 6.48 & 0.00 & 0.02 \\
N4325 & 47 & 904$\pm$59 & -13.7 & 3.0$\pm$2.1 & 0.22 & 0.26 & 0.01 & 0.02 &  & 0.25 & 2.14 & 0.00 & 0.02 \\
N4325 & 150 & 8151$\pm$40 & -17.6 & 13.2$\pm$0.9 & 0.12 & 1.39 & 0.07 & 0.09 &  & 0.05 & 1.30 & 0.05 & 0.05 \\
N4325 & 153 & 8256$\pm$34 & -17.4 & 8.9$\pm$1.1 & 0.00 & 2.07 & 0.08 & 0.08 &  & 0.00 & 1.92 & 0.06 & 0.07 \\
N4325 & 174 & 8146$\pm$40 & -16.9 & 13.0$\pm$1.7 & 0.00 & 1.59 & 0.01 & 0.03 &  & 0.01 & 1.45 & 0.03 & -0.00 \\
\tablenotetext{a}{$H_0=100$ \kms Mpc$^{-1}$, $q_0=0.5$, flat cosmology}
\tablenotetext{b}{Average error for $B/T$ measurement is $\sim0.1$ 
(Simard $et~al.$ 2000), although
$B/T$ can change by as much as 0.59 when the original ($cz_0$) and
degraded ($cz=9000$ \kms) values are compared for individual galaxies.}
\tablenotetext{c}{Average error for $r_{1/2}$ is $\sim$10\% (Simard
$et~al.$ 2000).}
\tablenotetext{d}{Average random error for $R_A$ is $\sim0.02$
(Simard $et~al.$ 2000; Im $et~al.$ 2000), although $R_A$ can change 
by as much as 0.16 when the original ($cz_0$) and degraded ($cz=9000$ \kms) 
values are compared.}
\tablenotetext{e}{Average random error for $R_T$ is $\sim0.02$ 
(Simard $et~al.$ 2000; Im $et~al.$ 2000), although $R_T$ can change by as 
much as 0.30 when the original ($cz_0$) and degraded ($cz=9000$ \kms) 
values are compared.}
\enddata
\end{deluxetable}

\begin{deluxetable}{lrrrrrrrrcrrrr}
\tabletypesize{\tiny}
\tablecolumns{14}
\tablewidth{0pc}
\tablecaption{Structural Parameters of Group Galaxies Measured at $cz_0$ 
and $cz=9000$ \kms \label{table3}}
\tablehead{
\multicolumn{5}{c}{}    & \multicolumn{4}{c}{$cz_0$}    &
\colhead{}          & \multicolumn{4}{c}{$cz=9000$ \kms} \\
\cline{6-9} \cline{11-14}   \\
\colhead{Field}     & \colhead{ID}  & \colhead{$cz_0$ (\kms)} &
\colhead{$M_R-5$log$h$\tablenotemark{a}}     & 
\colhead{[OII] EW (\AA)}      &
\colhead{B/T\tablenotemark{b}}
 &
\colhead{$r_{1/2}$\tablenotemark{c}~ (\hi kpc)}   & 
\colhead{$R_A$\tablenotemark{d}}   &
\colhead{$R_T$\tablenotemark{e}} &
\colhead{} &            \colhead{B/T} &
\colhead{$r_{1/2}$ (\hi kpc)}   & \colhead{$R_A$}   & \colhead{$R_T$} }
\startdata
H42 & 1 & 3950$\pm$30 & -21.5 & -0.4$\pm$0.7 & 0.73 & 6.49 & 0.02 & 0.03 &  & 0.84 & 6.43 & 0.03 & 0.08 \\
  & 4 & 4176$\pm$25 & -20.0 & -0.5$\pm$0.7 & 1.00 & 2.75 & 0.02 & 0.13 &  & 0.56 & 2.16 & 0.01 & 0.04 \\
  & 5 & 3980$\pm$23 & -20.1 & -0.0$\pm$0.6 & 1.00 & 1.44 & 0.05 & 0.17 &  & 0.74 & 1.69 & 0.02 & 0.06 \\
  & 6 & 3538$\pm$46 & -18.6 & 2.8$\pm$1.5 & 0.05 & 3.70 & 0.06 & 0.12 &  & 0.06 & 3.99 & 0.03 & 0.09 \\
  & 7 & 4193$\pm$25 & -19.8 & -0.5$\pm$0.8 & 0.71 & 0.91 & 0.02 & 0.07 &  & 0.71 & 0.77 & 0.01 & 0.02 \\
  & 9 & 3424$\pm$37 & -18.6 & 5.7$\pm$1.5 & 0.57 & 1.04 & 0.02 & 0.02 &  & 0.52 & 1.11 & 0.02 & 0.04 \\
  & 13 & 3504$\pm$36 & -18.5 & 10.2$\pm$1.1 & 0.03 & 3.37 & 0.02 & 0.16 &  & 0.03 & 3.43 & 0.01 & 0.15 \\
  & 14 & 3636$\pm$19 & -18.9 & -0.5$\pm$0.6 & 0.59 & 1.35 & 0.02 & 0.09 &  & 0.43 & 0.88 & 0.10 & 0.23 \\
  & 15 & 4287$\pm$31 & -18.2 & 22.1$\pm$1.5 & 0.30 & 3.42 & 0.08 & 0.13 &  & 0.22 & 3.25 & 0.05 & 0.10 \\
  & 21 & 3675$\pm$29 & -18.4 & 1.4$\pm$1.0 & 0.66 & 1.16 & 0.03 & 0.05 &  & 0.46 & 0.99 & 0.04 & 0.07 \\
  & 22 & 3828$\pm$26 & -17.9 & 0.8$\pm$1.0 & 0.31 & 1.48 & 0.01 & 0.03 &  & 0.22 & 1.54 & -0.01 & -0.00 \\
  & 23 & 3613$\pm$27 & -17.9 & 6.7$\pm$1.1 & 0.44 & 1.56 & 0.01 & 0.04 &  & 0.52 & 1.55 & -0.00 & 0.00 \\
  & 24 & 4076$\pm$80 & -18.3 & 0.0$\pm$0.0 & 0.46 & 1.05 & 0.01 & 0.02 &  & 0.26 & 0.85 & 0.00 & 0.00 \\
  & 28 & 3621$\pm$23 & -17.8 & -0.3$\pm$0.7 & 0.53 & 1.04 & 0.01 & 0.03 &  & 0.46 & 1.23 & 0.00 & 0.03 \\
  & 29 & 3766$\pm$24 & -17.6 & 0.2$\pm$0.6 & 0.46 & 0.98 & 0.01 & 0.03 &  & 0.41 & 0.99 & -0.00 & -0.00 \\
  & 34 & 3876$\pm$39 & -17.4 & 4.9$\pm$1.8 & 0.40 & 0.99 & 0.01 & 0.05 &  & 0.45 & 1.05 & 0.00 & -0.00 \\
  & 46 & 3891$\pm$39 & -16.9 & -1.0$\pm$0.5 & 0.79 & 1.37 & 0.02 & 0.03 &  & 0.38 & 1.47 & 0.02 & 0.02 \\
  & 59 & 3647$\pm$59 & -16.1 & 1.7$\pm$1.5 & 0.13 & 0.71 & 0.04 & 0.05 &  & 0.29 & 0.97 & 0.02 & 0.03 \\
  & 65 & 3938$\pm$72 & -16.4 & 5.0$\pm$2.5 & 0.52 & 0.97 & 0.02 & 0.04 &  & 0.22 & 0.97 & 0.01 & 0.01 \\
  & 69 & 3675$\pm$27 & -16.8 & -0.4$\pm$0.6 & 0.50 & 0.37 & 0.01 & 0.03 &  & 0.55 & 0.41 & -0.00 & 0.00 \\
  & 85 & 3402$\pm$32 & -15.7 & 7.4$\pm$1.2 & 0.00 & 0.74 & 0.00 & 0.04 &  & 0.51 & 1.24 & -0.01 & -0.01 \\
  & 94 & 3732$\pm$75 & -15.6 & 6.8$\pm$2.1 & 0.00 & 1.02 & 0.00 & 0.03 &  & 0.39 & 2.00 & -0.03 & 0.08 \\
  & 106 & 3924$\pm$132 & -15.7 & 0.1$\pm$0.9 & 0.09 & 0.82 & 0.03 & 0.01 &  & 0.68 & 1.22 & -0.07 & -0.03 \\
  & 136 & 4587$\pm$36 & -15.7 & 57.9$\pm$16.1 & 0.15 & 0.99 & 0.00 & 0.05 &  & 0.43 & 1.03 & -0.02 & -0.02 \\
  & 143 & 4081$\pm$43 & -15.5 & 0.3$\pm$1.1 & 0.14 & 0.67 & 0.00 & 0.04 &  & 0.33 & 0.68 & 0.02 & -0.01 \\
H62 & 1 & 4284$\pm$31 & -20.6 & 0.4$\pm$1.1 & 0.81 & 6.70 & 0.19 & 0.18 &  & 0.59 & 4.01 & 0.25 & 0.25 \\
  & 3 & 3690$\pm$27 & -19.9 & 3.1$\pm$1.2 & 0.74 & 3.68 & 0.04 & 0.14 &  & 0.83 & 4.64 & 0.05 & 0.17 \\
  & 4 & 3555$\pm$28 & -19.7 & -0.3$\pm$0.7 & 0.80 & 1.17 & 0.04 & 0.25 &  & 0.45 & 3.40 & 0.05 & 0.07 \\
  & 8 & 3565$\pm$14 & -20.5 & 11.3$\pm$1.8 & 0.83 & 3.01 & 0.03 & 0.08 &  & 0.81 & 4.74 & 0.02 & 0.08 \\
 & 9 & 4378$\pm$19 & -20.0 & -0.2$\pm$0.7 & 0.70 & 2.95 & 0.02 & 0.07 &  & 0.34 & 2.49 & 0.00 & 0.04 \\
  & 10 & 4382$\pm$28 & -20.3 & -0.0$\pm$0.8 & 0.75 & 2.10 & 0.03 & ... &  & 0.61 & 3.34 & -0.10 & -0.07 \\
  & 14 & 4166$\pm$28 & -19.9 & -0.6$\pm$0.8 & 0.59 & 1.84 & 0.04 & 0.16 &  & 0.92 & 2.10 & 0.03 & 0.07 \\
  & 16 & 4310$\pm$24 & -19.4 & -0.5$\pm$0.6 & 0.65 & 2.41 & 0.02 & 0.08 &  & 0.45 & 1.97 & 0.01 & 0.05 \\
  & 18 & 4424$\pm$17 & -19.5 & -0.2$\pm$1.0 & 0.89 & 6.01 & 0.27 & 0.41 &  & 0.42 & 5.12 & 0.11 & 0.11 \\
  & 19 & 4221$\pm$33 & -19.3 & 0.6$\pm$1.9 & 0.86 & 1.30 & 0.02 & 0.06 &  & 0.65 & 1.24 & 0.01 & 0.04 \\
  & 22 & 4871$\pm$26 & -19.6 & 0.1$\pm$0.8 & 0.79 & 2.08 & 0.02 & 0.06 &  & 0.58 & 1.66 & 0.01 & 0.03 \\
  & 25 & 4391$\pm$28 & -18.9 & -0.8$\pm$0.8 & 0.58 & 1.58 & 0.01 & 0.05 &  & 0.85 & 1.71 & -0.00 & 0.04 \\
  & 27 & 4196$\pm$17 & -18.8 & -0.4$\pm$0.7 & 0.56 & 1.07 & 0.02 & 0.04 &  & 0.90 & 1.16 & 0.02 & 0.05 \\
  & 29 & 4238$\pm$34 & -18.7 & -0.1$\pm$1.3 & 0.81 & 1.92 & 0.01 & 0.03 &  & 0.66 & 1.56 & 0.01 & 0.02 \\
  & 33 & 4076$\pm$30 & -18.4 & -0.5$\pm$0.9 & 0.80 & 0.91 & 0.01 & 0.03 &  & 0.81 & 0.89 & 0.00 & 0.01 \\
  & 36 & 4829$\pm$36 & -18.4 & -0.1$\pm$1.0 & 0.47 & 1.46 & 0.01 & 0.04 &  & 0.44 & 1.26 & 0.00 & 0.01 \\
  & 38 & 3962$\pm$33 & -17.7 & -0.7$\pm$0.7 & 0.25 & 1.21 & 0.01 & 0.01 &  & 0.27 & 1.31 & 0.00 & 0.00 \\
  & 39 & 4316$\pm$34 & -17.8 & 1.4$\pm$1.3 & 0.33 & 1.28 & 0.01 & 0.02 &  & 0.39 & 1.34 & 0.00 & 0.00 \\
  & 41 & 4700$\pm$20 & -17.5 & 48.1$\pm$5.3 & 0.30 & 2.15 & 0.09 & 0.10 &  & 0.21 & 1.87 & 0.07 & 0.07 \\
  & 43 & 3964$\pm$34 & -17.7 & -0.3$\pm$0.6 & 0.71 & 1.67 & 0.04 & 0.07 &  & 0.68 & 1.82 & 0.05 & 0.05 \\
  & 46 & 4731$\pm$34 & -17.1 & 17.0$\pm$1.6 & 0.11 & 1.95 & -0.00 & 0.05 &  & 0.10 & 1.78 & 0.01 & 0.01 \\
  & 55 & 4655$\pm$50 & -17.6 & 6.4$\pm$2.0 & 0.28 & 1.36 & 0.00 & 0.01 &  & 0.20 & 1.25 & 0.01 & -0.00 \\
  & 56 & 4849$\pm$42 & -17.5 & 1.3$\pm$1.1 & 0.34 & 1.46 & 0.00 & 0.03 &  & 0.37 & 1.38 & -0.00 & 0.02 \\
  & 61 & 4695$\pm$37 & -16.8 & 11.7$\pm$2.5 & 0.01 & 2.32 & 0.06 & 0.07 &  & 0.01 & 2.13 & 0.04 & 0.03 \\
  & 62 & 4843$\pm$34 & -17.4 & -0.6$\pm$0.5 & 0.15 & 0.78 & 0.03 & 0.04 &  & 0.01 & 0.88 & 0.15 & 0.15 \\
  & 66 & 4826$\pm$44 & -17.5 & 1.3$\pm$1.3 & 0.51 & 1.06 & 0.04 & 0.05 &  & 0.36 & 0.82 & 0.01 & 0.02 \\
  & 75 & 4397$\pm$119 & -16.8 & 11.5$\pm$1.9 & 0.01 & 1.39 & 0.02 & 0.03 &  & 0.01 & 1.38 & -0.01 & 0.01 \\
  & 76 & 4963$\pm$35 & -16.7 & 10.4$\pm$0.9 & 0.52 & 0.84 & 0.01 & 0.02 &  & 0.75 & 0.84 & -0.01 & -0.01 \\
  & 89 & 4208$\pm$78 & -16.6 & 6.6$\pm$2.4 & 0.20 & 0.90 & 0.09 & 0.13 &  & 0.55 & 1.28 & 0.02 & 0.01 \\
  & 100 & 3917$\pm$102 & -15.7 & 2.9$\pm$1.6 & 0.01 & 2.30 & 0.01 & 0.03 &  & 0.01 & 2.44 & -0.02 & -0.01 \\
  & 109 & 4688$\pm$43 & -16.1 & 36.5$\pm$8.3 & 0.29 & 1.46 & 0.02 & 0.03 &  & 0.43 & 1.64 & 0.01 & -0.03 \\
  & 111 & 3485$\pm$103 & -16.2 & 3.7$\pm$3.0 & 0.01 & 0.56 & 0.02 & 0.04 &  & 0.02 & 0.68 & 0.00 & 0.02 \\
  & 121 & 4235$\pm$103 & -16.2 & 1.5$\pm$1.2 & 0.28 & 0.72 & -0.00 & 0.01 &  & 0.52 & 0.73 & 0.01 & 0.00 \\
  & 132 & 4285$\pm$75 & -16.0 & -2.4$\pm$2.6 & 0.51 & 0.62 & 0.01 & 0.01 &  & 0.43 & 0.77 & 0.00 & 0.01 \\
  & 141 & 3722$\pm$44 & -15.8 & -1.8$\pm$0.8 & 0.20 & 0.73 & -0.01 & 0.00 &  & 0.23 & 0.83 & -0.00 & -0.00 \\
  & 150 & 4671$\pm$124 & -16.3 & -9.8$\pm$-9.8 & 0.46 & 1.00 & 0.00 & 0.01 &  & 0.33 & 0.86 & -0.02 & -0.01 \\
  & 158 & 3953$\pm$71 & -15.9 & 1.0$\pm$1.2 & 0.13 & 0.93 & 0.01 & 0.02 &  & 0.34 & 1.03 & -0.01 & -0.01 \\
  & 160 & 4228$\pm$63 & -15.7 & 8.1$\pm$3.7 & 0.09 & 0.97 & 0.02 & 0.01 &  & 0.19 & 1.12 & -0.01 & 0.00 \\
N2563 & 1 & 4658$\pm$42 & -21.1 & 1.8$\pm$1.6 & 0.94 & 6.28 & 0.01 & 0.04 &  & 0.88 & 6.80 & 0.02 & 0.04 \\
  & 3 & 4857$\pm$45 & -20.7 & -0.0$\pm$1.0 & 1.00 & 3.95 & 0.01 & 0.11 &  & 0.82 & 3.79 & 0.02 & 0.12 \\
  & 5 & 4812$\pm$28 & -20.6 & 0.7$\pm$0.9 & 0.68 & 4.45 & 0.03 & 0.16 &  & 0.49 & 3.61 & 0.01 & 0.06 \\
  & 8 & 5267$\pm$30 & -21.1 & 0.6$\pm$1.6 & 0.64 & 2.67 & 0.04 & 0.07 &  & 0.52 & 2.59 & 0.04 & 0.06 \\
  & 9 & 4110$\pm$40 & -18.4 & 3.8$\pm$0.7 & 0.02 & 3.00 & 0.06 & -0.14 &  & 0.09 & 4.40 & 0.06 & 0.10 \\
  & 10 & 5567$\pm$10 & -20.2 & 18.6$\pm$3.4 & 0.54 & 5.19 & 0.15 & 0.24 &  & 0.52 & 5.10 & 0.09 & 0.15 \\
  & 13 & 5148$\pm$24 & -19.9 & -0.3$\pm$0.8 & 0.91 & 2.20 & 0.02 & 0.06 &  & 0.84 & 1.75 & 0.04 & 0.06 \\
  & 16 & 5346$\pm$68 & -19.4 & 2.7$\pm$2.8 & 0.16 & 2.71 & 0.02 & 0.07 &  & 0.11 & 2.36 & 0.03 & 0.08 \\
  & 18 & 4114$\pm$32 & -18.9 & 2.6$\pm$0.8 & 0.07 & 3.84 & 0.04 & 0.17 &  & 0.03 & 4.42 & 0.03 & 0.13 \\
  & 19 & 4529$\pm$29 & -19.3 & 2.7$\pm$3.0 & 0.52 & 1.12 & 0.04 & 0.17 &  & 0.54 & 1.07 & 0.01 & 0.02 \\
  & 21 & 4594$\pm$30 & -19.2 & -1.2$\pm$0.7 & 0.70 & 1.08 & 0.03 & 0.09 &  & 0.86 & 1.27 & 0.02 & 0.04 \\
  & 23 & 4520$\pm$14 & -18.6 & 11.9$\pm$1.4 & 0.01 & 2.42 & 0.10 & 0.11 &  & 0.00 & 2.44 & 0.07 & 0.08 \\
  & 24 & 4914$\pm$10 & -18.8 & 7.7$\pm$1.1 & 0.41 & 1.74 & 0.05 & 0.08 &  & 0.11 & 1.47 & 0.04 & 0.04 \\
  & 27 & 5040$\pm$36 & -18.7 & 1.8$\pm$0.7 & 0.13 & 5.65 & 0.03 & 0.11 &  & 0.05 & 4.78 & 0.02 & -0.03 \\
  & 28 & 5307$\pm$78 & -18.8 & 2.3$\pm$1.6 & 0.10 & 3.33 & 0.02 & 0.06 &  & 0.04 & 2.97 & 0.02 & 0.03 \\
  & 30 & 4674$\pm$27 & -18.7 & -0.2$\pm$0.6 & 0.81 & 1.76 & 0.01 & 0.03 &  & 0.70 & 1.84 & 0.01 & 0.05 \\
  & 33 & 4270$\pm$24 & -17.8 & 17.6$\pm$3.2 & 0.19 & 2.11 & 0.04 & 0.08 &  & 0.15 & 2.64 & 0.09 & 0.12 \\
  & 34 & 4661$\pm$50 & -18.6 & -0.3$\pm$0.9 & 0.71 & 0.49 & 0.02 & 0.05 &  & 0.58 & 0.76 & 0.01 & 0.01 \\
  & 35 & 5061$\pm$116 & -18.2 & 5.8$\pm$8.5 & 0.08 & 2.23 & 0.02 & 0.08 &  & 0.00 & 1.91 & 0.04 & 0.05 \\
  & 38 & 3945$\pm$31 & -17.7 & 0.0$\pm$0.9 & 0.72 & 2.50 & 0.01 & 0.04 &  & 0.52 & 2.67 & 0.01 & 0.01 \\
  & 40 & 4640$\pm$72 & -18.0 & 5.7$\pm$2.9 & 0.10 & 1.68 & 0.01 & 0.04 &  & 0.17 & 1.80 & 0.00 & 0.02 \\
  & 41 & 4917$\pm$25 & -18.7 & -0.1$\pm$0.6 & 0.54 & 1.25 & 0.01 & 0.04 &  & 0.59 & 0.90 & 0.01 & 0.03 \\
  & 43 & 4473$\pm$80 & -18.0 & 2.6$\pm$3.7 & 0.23 & 1.12 & 0.01 & 0.03 &  & 0.32 & 1.37 & 0.01 & 0.02 \\
  & 44 & 4920$\pm$57 & -17.8 & -0.7$\pm$0.7 & 0.06 & 1.10 & 0.00 & 0.02 &  & 0.07 & 1.08 & -0.00 & -0.01 \\
  & 50 & 4788$\pm$57 & -17.7 & 5.3$\pm$4.2 & 1.00 & 1.62 & 0.01 & 0.06 &  & 0.71 & 1.33 & 0.00 & 0.02 \\
  & 51 & 4606$\pm$73 & -17.7 & -1.1$\pm$1.2 & 0.68 & 0.60 & 0.00 & -0.00 &  & 0.90 & 0.84 & -0.01 & 0.05 \\
  & 53 & 4286$\pm$39 & -17.4 & 6.0$\pm$1.3 & 0.00 & 1.47 & 0.02 & 0.08 &  & 0.00 & 1.64 & 0.02 & 0.02 \\
  & 58 & 4868$\pm$65 & -17.5 & 2.8$\pm$10.2 & 0.36 & 1.76 & 0.01 & 0.05 &  & 0.21 & 1.77 & -0.00 & 0.00 \\
  & 82 & 5176$\pm$110 & -17.3 & 8.8$\pm$3.2 & 0.77 & 1.11 & 0.01 & 0.03 &  & 0.56 & 0.92 & 0.00 & 0.01 \\
  & 92 & 4504$\pm$45 & -17.1 & 1.3$\pm$1.5 & 0.41 & 1.80 & 0.01 & 0.03 &  & 0.28 & 2.02 & -0.00 & 0.05 \\
  & 94 & 4844$\pm$37 & -16.9 & 21.3$\pm$1.4 & 0.14 & 0.88 & 0.02 & 0.04 &  & 0.43 & 1.09 & -0.01 & 0.00 \\
  & 114 & 5076$\pm$87 & -16.8 & 11.3$\pm$4.5 & 0.53 & 0.95 & -0.01 & 0.01 &  & 0.80 & 1.30 & -0.00 & 0.00 \\
  & 120 & 4870$\pm$84 & -16.7 & 3.5$\pm$2.3 & 0.58 & 1.10 & 0.01 & 0.03 &  & 0.53 & 1.17 & -0.01 & 0.04 \\
  & 148 & 4055$\pm$44 & -16.1 & -0.3$\pm$0.9 & 0.40 & 0.84 & 0.01 & 0.05 &  & 0.46 & 0.91 & -0.01 & -0.01 \\
  & 150 & 4557$\pm$37 & -16.2 & 41.9$\pm$5.3 & 0.01 & 1.28 & 0.02 & 0.08 &  & 0.03 & 1.41 & 0.01 & 0.07 \\
  & 163 & 4375$\pm$77 & -15.8 & 0.7$\pm$2.5 & 0.24 & 0.85 & 0.03 & 0.11 &  & 0.44 & 0.94 & 0.01 & -0.00 \\
N3557 & 1 & 3009$\pm$27 & -21.5 & -0.2$\pm$0.7 & 0.75 & 5.54 & 0.04 & -0.04 &  & 0.93 & 6.27 & 0.05 & 0.10 \\
  & 5 & 2414$\pm$33 & -18.9 & 7.0$\pm$0.8 & 0.46 & 5.71 & 0.16 & ... &  & 0.14 & 13.44 & -0.14 & ... \\
  & 6 & 3469$\pm$26 & -19.6 & -0.4$\pm$0.7 & 0.42 & 3.17 & 0.05 & 0.12 &  & 0.23 & 2.44 & 0.01 & 0.06 \\
  & 7 & 3008$\pm$18 & -20.2 & -0.4$\pm$0.7 & 0.57 & 1.12 & 0.02 & 0.09 &  & 0.62 & 1.09 & 0.02 & 0.06 \\
  & 9 & 2784$\pm$26 & -20.2 & -0.5$\pm$0.9 & 0.67 & 2.27 & 0.02 & 0.09 &  & 0.69 & 2.48 & 0.07 & 0.11 \\
  & 11 & 2782$\pm$31 & -17.7 & -0.5$\pm$0.7 & 0.11 & 1.77 & 0.02 & 0.07 &  & 0.27 & 1.99 & 0.03 & 0.04 \\
  & 16 & 3183$\pm$41 & -16.5 & 15.8$\pm$1.7 & 0.04 & 2.00 & -0.01 & 0.04 &  & 0.02 & 2.18 & 0.02 & 0.05 \\
  & 19 & 3080$\pm$40 & -16.8 & 1.5$\pm$1.6 & 0.14 & 1.16 & 0.01 & 0.03 &  & 0.64 & 1.74 & -0.03 & 0.01 \\
  & 25 & 2772$\pm$44 & -15.8 & 8.8$\pm$2.0 & 0.00 & 1.42 & 0.02 & 0.07 &  & 0.01 & 1.30 & 0.01 & 0.01 \\
  & 32 & 2623$\pm$65 & -15.1 & 3.8$\pm$1.9 & 0.00 & 1.32 & 0.01 & 0.04 &  & 0.08 & 1.45 & -0.00 & 0.06 \\
  & 38 & 3062$\pm$71 & -15.7 & 0.2$\pm$3.6 & 0.09 & 1.15 & -0.00 & 0.02 &  & 0.15 & 1.34 & -0.00 & 0.00 \\
  & 47 & 3146$\pm$36 & -15.8 & 47.2$\pm$1.5 & 0.57 & 0.44 & 0.04 & 0.06 &  & 0.63 & 0.37 & -0.00 & -0.00 \\
  & 49 & 2640$\pm$68 & -14.8 & 1.2$\pm$1.2 & 0.02 & 0.61 & 0.01 & 0.02 &  & 0.50 & 0.84 & -0.03 & 0.03 \\
  & 97 & 2751$\pm$102 & -14.7 & 0.1$\pm$0.8 & 0.10 & 0.72 & -0.00 & 0.01 &  & 0.55 & 0.86 & 0.04 & -0.01 \\
  & 106 & 3300$\pm$132 & -14.6 & -0.0$\pm$1.1 & 0.02 & 0.74 & 0.05 & 0.04 &  & 0.56 & 1.03 & -0.07 & -0.00 \\
N4325 & 4 & 7564$\pm$41 & -21.0 & 5.4$\pm$2.0 & 0.81 & 6.73 & 0.02 & 0.06 &  & 0.77 & 6.49 & 0.01 & 0.06 \\
  & 7 & 7779$\pm$30 & -20.8 & 3.3$\pm$1.4 & 0.79 & 2.96 & 0.08 & 0.15 &  & 0.46 & 2.54 & 0.14 & 0.29 \\
  & 8 & 7941$\pm$27 & -20.6 & 0.8$\pm$1.3 & 0.83 & 3.97 & 0.02 & 0.12 &  & 0.41 & 3.09 & 0.02 & 0.00 \\
  & 11 & 7747$\pm$15 & -19.8 & 4.9$\pm$1.4 & 0.10 & 3.31 & 0.04 & 0.09 &  & 0.08 & 2.95 & 0.03 & 0.07 \\
  & 16 & 7661$\pm$36 & -19.2 & 3.5$\pm$1.1 & 0.01 & 2.25 & 0.04 & 0.07 &  & 0.00 & 2.14 & 0.02 & 0.06 \\
  & 17 & 7905$\pm$26 & -19.2 & 0.9$\pm$1.0 & 0.25 & 1.68 & 0.03 & 0.06 &  & 0.27 & 1.57 & 0.01 & 0.04 \\
  & 29 & 7187$\pm$23 & -18.8 & 1.0$\pm$1.2 & 0.49 & 1.10 & 0.01 & 0.03 &  & 0.47 & 1.27 & 0.01 & 0.01 \\
  & 30 & 7316$\pm$37 & -18.8 & 21.7$\pm$0.7 & 0.40 & 0.93 & 0.04 & 0.07 &  & 0.46 & 0.95 & 0.01 & 0.02 \\
  & 35 & 7397$\pm$60 & -18.6 & 12.1$\pm$2.6 & 0.71 & 1.45 & 0.04 & 0.06 &  & 0.57 & 1.37 & 0.03 & 0.04 \\
  & 37 & 7340$\pm$28 & -18.5 & -0.8$\pm$0.6 & 0.70 & 1.49 & 0.02 & 0.03 &  & 0.60 & 1.51 & 0.00 & 0.01 \\
  & 50 & 7416$\pm$42 & -18.1 & 1.8$\pm$1.8 & 0.51 & 0.73 & 0.01 & 0.03 &  & 0.63 & 0.84 & 0.00 & 0.01 \\
  & 55 & 7018$\pm$49 & -17.9 & 0.8$\pm$2.0 & 0.54 & 1.61 & 0.02 & 0.04 &  & 0.47 & 1.66 & 0.01 & 0.01 \\
  & 60 & 7419$\pm$45 & -18.0 & 22.4$\pm$2.7 & 0.00 & 1.82 & 0.10 & 0.15 &  & 0.00 & 1.80 & 0.04 & 0.09 \\
  & 78 & 7650$\pm$67 & -18.0 & -9.7$\pm$-9.7 & 0.05 & 2.50 & 0.01 & 0.06 &  & 0.15 & 2.46 & 0.00 & 0.06 \\
  & 79 & 7696$\pm$116 & -18.0 & 4.6$\pm$1.2 & 0.12 & 2.01 & 0.02 & 0.05 &  & 0.17 & 2.03 & 0.00 & 0.00 \\
  & 81 & 7707$\pm$80 & -17.9 & 3.5$\pm$1.9 & 0.11 & 0.94 & 0.01 & 0.02 &  & 0.15 & 0.95 & -0.00 & 0.00 \\
  & 83 & 7238$\pm$32 & -17.6 & 12.6$\pm$0.9 & 0.35 & 0.98 & 0.03 & 0.04 &  & 0.33 & 1.04 & 0.01 & 0.03 \\
  & 89 & 7972$\pm$62 & -17.7 & -0.3$\pm$1.1 & 0.63 & 1.24 & 0.03 & 0.05 &  & 0.39 & 1.02 & 0.00 & 0.02 \\
  & 106 & 7647$\pm$38 & -17.5 & 3.1$\pm$1.8 & 0.23 & 0.72 & 0.06 & 0.05 &  & 0.32 & 0.72 & 0.02 & 0.03 \\
  & 129 & 7696$\pm$42 & -17.1 & 0.7$\pm$1.1 & 0.48 & 1.01 & 0.01 & 0.03 &  & 0.47 & 1.13 & -0.00 & 0.00 \\
  & 133 & 7656$\pm$61 & -17.5 & 5.0$\pm$2.1 & 0.64 & 1.02 & 0.01 & 0.02 &  & 0.63 & 1.01 & 0.00 & 0.01 \\
  & 136 & 6978$\pm$41 & -17.4 & 14.4$\pm$1.5 & 0.00 & 1.71 & 0.07 & 0.09 &  & 0.00 & 1.85 & 0.05 & 0.06 \\
  & 137 & 6489$\pm$40 & -16.8 & 23.4$\pm$1.1 & 0.34 & 0.99 & 0.01 & 0.04 &  & 0.35 & 1.14 & 0.01 & -0.00 \\
N5129 & 1 & 6956$\pm$30 & -21.9 & 1.7$\pm$1.7 & 0.90 & 9.10 & 0.02 & 0.05 &  & 0.89 & 9.03 & 0.04 & 0.09 \\
  & 2 & 7285$\pm$25 & -21.4 & 0.3$\pm$0.8 & 0.75 & 7.31 & 0.03 & 0.15 &  & 0.73 & 6.99 & 0.03 & 0.16 \\
  & 4 & 6817$\pm$37 & -20.8 & 1.6$\pm$0.9 & 0.88 & 4.32 & 0.02 & 0.04 &  & 0.77 & 4.00 & 0.02 & 0.07 \\
  & 5 & 7239$\pm$14 & -20.4 & 8.6$\pm$1.4 & 0.20 & 6.14 & 0.11 & 0.19 &  & 0.05 & 5.06 & 0.09 & 0.11 \\
  & 6 & 6433$\pm$59 & -19.8 & 3.9$\pm$1.3 & 0.05 & 4.92 & 0.07 & ... &  & 0.07 & 10.16 & ... & ... \\
  & 7 & 6839$\pm$25 & -19.9 & 4.6$\pm$2.3 & 0.29 & 3.20 & 0.02 & 0.04 &  & 0.32 & 3.25 & 0.01 & 0.04 \\
  & 8 & 7031$\pm$51 & -19.7 & 1.8$\pm$1.8 & 0.10 & 4.78 & 0.03 & 0.07 &  & 0.02 & 4.54 & 0.03 & 0.08 \\
  & 9 & 7273$\pm$35 & -19.8 & 0.8$\pm$1.2 & 0.05 & 3.07 & 0.02 & 0.06 &  & 0.04 & 2.84 & 0.02 & 0.04 \\
  & 10 & 6441$\pm$18 & -19.4 & 8.9$\pm$1.3 & 0.06 & 3.34 & 0.05 & 0.08 &  & 0.04 & 3.60 & 0.05 & 0.07 \\
  & 11 & 7451$\pm$48 & -19.7 & 1.2$\pm$1.1 & 0.06 & 3.86 & 0.02 & 0.03 &  & 0.06 & 3.58 & 0.02 & 0.04 \\
  & 12 & 6754$\pm$55 & -19.5 & 3.5$\pm$1.1 & 0.41 & 1.96 & 0.03 & 0.06 &  & 0.48 & 2.16 & 0.02 & 0.03 \\
  & 13 & 7157$\pm$37 & -19.8 & -1.4$\pm$1.1 & 0.98 & 1.63 & 0.02 & 0.04 &  & 1.00 & 1.55 & 0.02 & 0.05 \\
  & 17 & 6768$\pm$81 & -19.0 & -0.4$\pm$0.7 & 0.48 & 2.52 & 0.03 & 0.04 &  & 0.46 & 2.43 & 0.03 & 0.03 \\
  & 21 & 7061$\pm$32 & -18.9 & 10.4$\pm$0.8 & 0.00 & 1.15 & 0.06 & 0.11 &  & 0.01 & 1.16 & 0.03 & 0.05 \\
  & 30 & 7026$\pm$44 & -18.5 & 2.2$\pm$1.3 & 0.37 & 1.42 & 0.01 & 0.01 &  & 0.49 & 1.38 & 0.02 & 0.01 \\
  & 35 & 7105$\pm$70 & -18.4 & 26.3$\pm$7.0 & 0.50 & 1.49 & 0.01 & 0.03 &  & 0.45 & 1.43 & 0.01 & 0.03 \\
  & 38 & 7132$\pm$40 & -18.4 & 3.6$\pm$1.7 & 0.14 & 1.37 & 0.01 & 0.02 &  & 0.18 & 1.35 & 0.01 & 0.01 \\
  & 39 & 7393$\pm$35 & -18.4 & -0.7$\pm$0.6 & 0.80 & 2.50 & -0.01 & 0.02 &  & 0.67 & 2.86 & -0.00 & 0.01 \\
  & 44 & 6905$\pm$36 & -18.4 & 6.4$\pm$2.3 & 0.20 & 2.85 & 0.02 & 0.06 &  & 0.06 & 2.67 & 0.02 & 0.05 \\
  & 52 & 7058$\pm$53 & -18.5 & 6.1$\pm$1.6 & 0.13 & 2.59 & 0.05 & 0.07 &  & 0.06 & 2.47 & 0.03 & 0.04 \\
  & 56 & 7261$\pm$72 & -18.1 & 5.8$\pm$1.7 & 0.00 & 1.88 & 0.01 & 0.03 &  & 0.00 & 1.86 & 0.00 & 0.03 \\
  & 58 & 6791$\pm$61 & -18.3 & 5.9$\pm$1.4 & 0.19 & 2.65 & 0.02 & 0.15 &  & 0.29 & 2.84 & 0.03 & 0.13 \\
  & 61 & 6890$\pm$55 & -18.0 & 8.9$\pm$2.6 & 0.07 & 1.57 & 0.01 & 0.03 &  & 0.05 & 1.56 & 0.00 & 0.02 \\
  & 62 & 7526$\pm$61 & -18.4 & 2.6$\pm$0.9 & 0.30 & 2.31 & -0.00 & 0.02 &  & 0.17 & 2.14 & 0.01 & 0.02 \\
  & 83 & 7239$\pm$82 & -18.0 & 6.1$\pm$1.5 & 0.00 & 2.04 & -0.00 & 0.04 &  & 0.00 & 1.94 & 0.01 & 0.04 \\
  & 95 & 7230$\pm$23 & -17.5 & 20.6$\pm$2.6 & 0.00 & 2.67 & 0.04 & 0.10 &  & 0.00 & 2.52 & 0.03 & 0.07 \\
  & 101 & 6233$\pm$45 & -17.3 & 14.2$\pm$2.4 & 0.01 & 1.49 & 0.02 & 0.04 &  & 0.00 & 1.61 & 0.01 & 0.02 \\
  & 102 & 6842$\pm$77 & -17.4 & 2.0$\pm$1.3 & 0.24 & 1.47 & 0.02 & 0.02 &  & 0.27 & 1.44 & -0.01 & 0.01 \\
  & 107 & 7081$\pm$38 & -17.5 & 0.8$\pm$1.0 & 0.66 & 1.14 & 0.02 & 0.03 &  & 0.75 & 1.12 & 0.01 & 0.02 \\
  & 113 & 7387$\pm$71 & -17.4 & -1.7$\pm$1.4 & 0.75 & 0.89 & 0.00 & 0.02 &  & 0.84 & 0.93 & 0.01 & 0.02 \\
  & 122 & 7060$\pm$66 & -17.1 & -0.3$\pm$1.4 & 0.26 & 1.61 & 0.01 & 0.03 &  & 0.23 & 1.61 & 0.00 & 0.01 \\
  & 145 & 6760$\pm$40 & -16.9 & 13.9$\pm$2.8 & 0.14 & 2.57 & 0.02 & 0.06 &  & 0.03 & 2.65 & 0.01 & 0.03 \\
  & 159 & 7242$\pm$86 & -17.2 & 2.5$\pm$2.2 & 0.03 & 0.97 & 0.01 & 0.03 &  & 0.01 & 0.94 & -0.00 & 0.00 \\
  & 9899 & 6701$\pm$39 & -16.7 & 18.8$\pm$2.4 & 0.16 & 0.85 & 0.02 & 0.04 &  & 0.26 & 0.90 & 0.02 & 0.01 \\
\tablenotetext{a}{$H_0=100$ \kms Mpc$^{-1}$, $q_0=0.5$, flat cosmology}
\tablenotetext{b}{Average error for $B/T$ measurement is $\sim0.1$ 
(Simard $et~al.$ 2000), although
$B/T$ can change by as much as 0.59 when the original ($cz_0$) and
degraded ($cz=9000$ \kms) values are compared for the individual galaxies.}
\tablenotetext{c}{Average error for $r_{1/2}$ is $\sim$10\% (Simard
$et~al.$ 2000).}
\tablenotetext{d}{Average random error for $R_A$ is $\sim0.02$ 
(Simard $et~al.$ 2000; Im $et~al.$ 2000), although $R_A$ can change by 
as much as 0.16 when the original ($cz_0$) and degraded ($cz=9000$ \kms) 
values are compared.}
\tablenotetext{e}{Average random error for $R_T$ is $\sim0.02$
(Simard $et~al.$ 2000; Im $et~al.$ 2000), although 
$R_T$ can change by as much as 0.30 when the original ($cz_0$) and
degraded ($cz=9000$ \kms) values are compared.}
\enddata
\end{deluxetable}

\begin{deluxetable}{lrrrcrr}
\tablecolumns{7}
\tablewidth{0pc}
\tablecaption{Global Group Characteristics \label{table4}}
\tablehead{
\multicolumn{2}{c}{}	& \multicolumn{2}{c}{$z_0$} & \colhead{} &
\multicolumn{2}{c}{$cz=9000$ \kms} \\
\cline{3-4}     \cline{6-7} \\

\colhead{Group}    &    \colhead{$\bar{v}$}   	&       
\colhead{(Early Types)$_T$\tablenotemark{a}} &	
\colhead{(Early Types)$_I$\tablenotemark{b}}&
\colhead{} &
\colhead{(Early Types)$_T$}&	\colhead{(Early Types)$_I$}	}
\startdata
HCG 42 	& 3830$\pm$47 & 15/25  (60\%) &  12/19  (63\%) & &  15/25  (60\%) &  13/19  (68\%) \\
HCG 62 	& 4356$\pm$50 & 21/38  (55\%) &  15/25  (60\%) & &  21/38  (55\%) &  15/25  (60\%) \\
NGC2563 & 4775$\pm$65 & 20/36  (56\%) &  14/20  (70\%) & &  20/36  (56\%) &  12/20  (60\%) \\
NGC3557 & 2843$\pm$64 & 6/15   (40\%) &  5/12   (42\%) & &  8/15   (53\%) &  7/12   (58\%) \\
NGC4325 & 7558$\pm$70 & 12/23  (52\%) &  8/11   (73\%) & &  11/23  (48\%) &  7/11   (64\%) \\
NGC5129 & 6998$\pm$51 & 10/34  (29\%) &  8/19   (42\%) & &  11/34  (32\%) &  8/19   (42\%) \\
\\
{\it Average} &       & 84/171 (49\%) &  62/106 (58\%) & &  86/171 (50\%) &  62/106 (58\%) \\
\tablenotetext{a}{The subscript $T$ denotes all group members in the 
$1^{\circ}\times1^{\circ}$ field.} 
\tablenotetext{b}{The subscript $I$ denotes only group members within 
$R\leq0.25$\hi Mpc (H$_0=100 h$ \kms Mpc$^{-1}$; q$_0=0.5$) of the group's center.}
\enddata
\end{deluxetable}

\begin{figure}
\caption{$\ref{fig1}a$:  Thumbnails of all the galaxies in
our sample (189) at their observed redshift.  The pixel area in each
thumbnail is 12 times the galaxy's isophotal area as defined by
SExtractor.  Reference numbers and absolute magnitudes corresponding
to Table 3 are listed in the bottom and upper left corners
respectively.  In these images, the seeing ranges from $1.''3$ to
$2.''6$.  $\ref{fig1}b$: The best-fit de Vaucouleurs bulge with
exponential disk models for the same galaxies found by GIM2D.  The
bulge fraction is listed in the upper left corner.  $\ref{fig1}c$:
The residual images created by subtracting the best GIM2D model from
the original galaxy image.  In the upper left corner, the asymmetry
parameter and total fraction of residual light are shown ($R_A, R_T$).
\label{fig1}}
\end{figure}
\begin{figure}
\end{figure}
\begin{figure}
\end{figure}

\begin{figure}
\caption{The same scheme as in Fig.~\ref{fig1}
but now for the redshifted ($cz=9000$\kms) galaxies.  The effective
resolution for all the galaxies is now 0.87\hi kpc.
\label{fig2}}
\end{figure}
\begin{figure}
\end{figure}
\begin{figure}
\end{figure}

\begin{figure}
\plotone{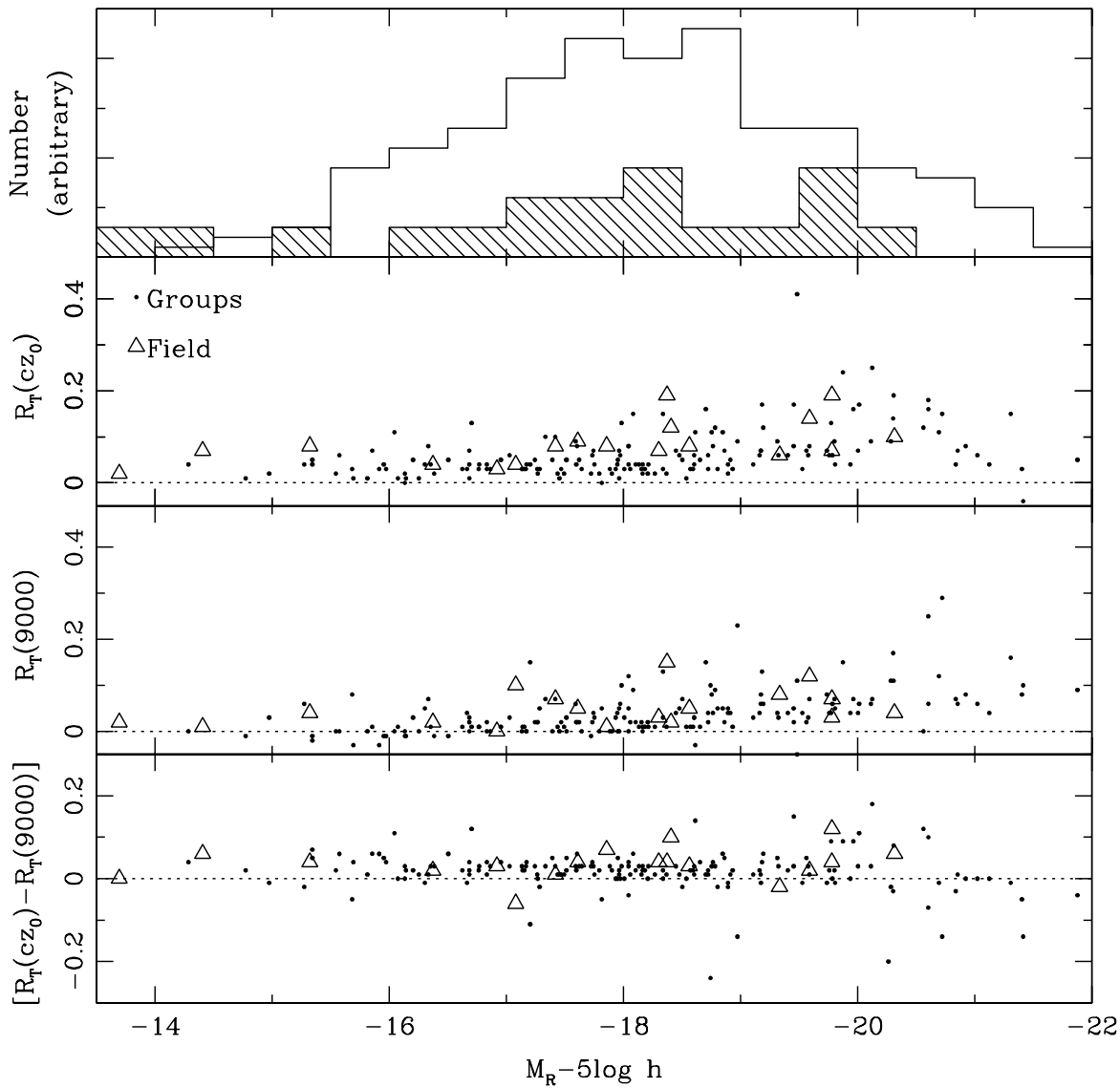}
\caption{{\it Top Panel:} Absolute magnitude
distribution of group (open bins) and field (hatched bins) samples;
the field sample is scaled arbitrarily by a factor of three to make
comparison of the samples easier.  Both samples roughly cover the same
absolute magnitude range, and a Kolmolgorov-Smirnov test finds the two
distributions are indistinguishable at the 95\% confidence level.
{\it Middle Panels:} The total fraction of residual light ($R_T$) as a
function of absolute magnitude ($M_R$) at the observed ($cz_0$) and
common ($cz=9000$\kms) redshift.  Field galaxies are shown as open
triangles and group galaxies as filled circles.  {\it Bottom Panel:}
The difference between the total residuals, $[R_T(cz_0) - R_T(9000)]$.
For galaxies fainter than $M_R=-20+5$log$h$, the fits ``improve''
slightly ($0.02\pm0.01$) after they are redshifted to $cz=9000$\kms.
\label{fig3}}
\end{figure}

\begin{figure}
\plotone{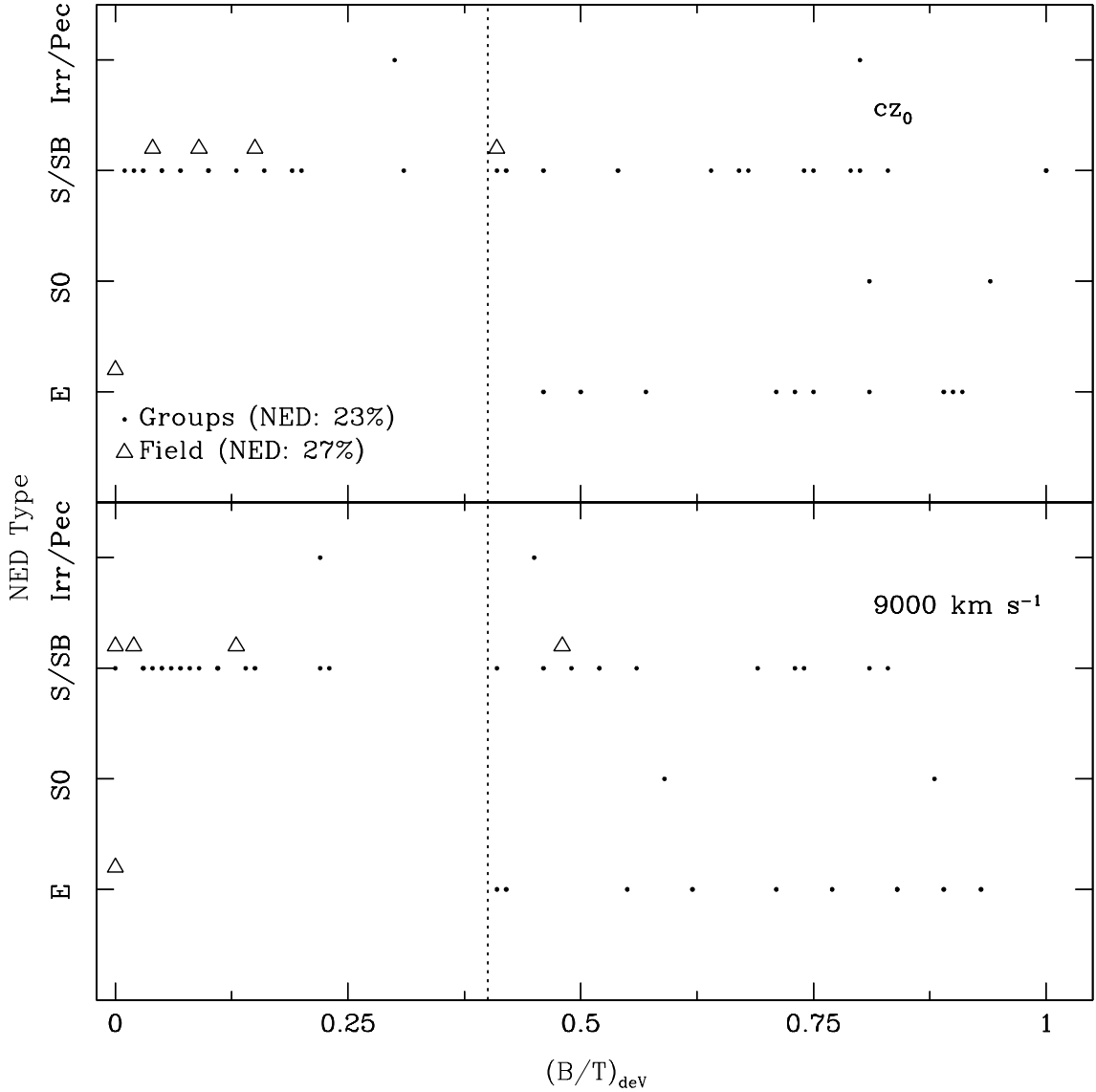}
\caption{Comparison of measured $B/T$ to published
morphological type.  We split the morphological types into four bins:
Ellipticals, S0's, Spirals/Barred Spirals, and Irregulars/Peculiars.
The upper panels correspond to the measured $B/T$ for each galaxy at
its observed redshift for the group and field.  The field points are
offset slightly in the y-direction for clarity.  Only a small set of
galaxies ($\sim24$\%) in our sample have been morphologically
classified in the literature.  Comparing type to our measured $B/T$,
we define ``early-type'' galaxies to have $B/T\ge 0.4$.
\label{fig4}}
\end{figure}

\begin{figure}
\plotone{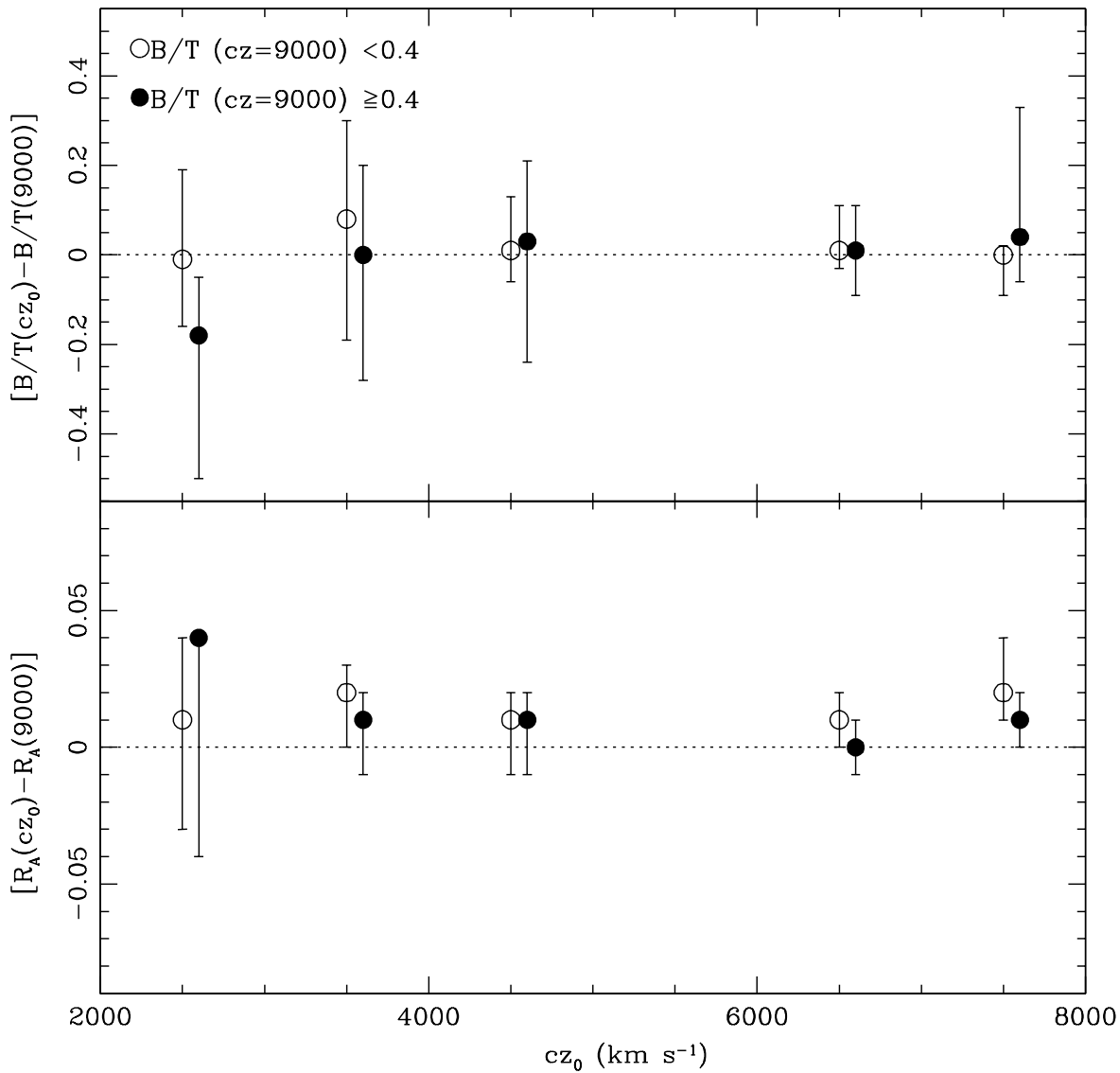}
\caption{{\it Top Panel:} The median change in
$B/T$ when the group galaxies are moved from $cz_0$ to $cz=9000$\kms~
and binned to the same effective resolution (0.87\hi kpc).  The
redshift bins are 1000\kms~wide, and we use the group's (not the
galaxy's) redshift.  The open circles are for galaxies with
$(B/T)_{cz_{9000}}<0.4$ and the filled circles for galaxies with
$(B/T)_{cz_{9000}}\geq0.4$; the two points in each bin are offset
slightly in the x-direction for clarity.  The asymmetric errorbars
correspond to $1\sigma$ in the $[(B/T)_{cz_0}-(B/T)_{cz_{9000}}]$
distribution about the median; lack of an errorbar indicates clumping
at that value.  Although galaxies have been redshifted by up to a
factor of four and the effective resolution degraded by as much as a
factor of five, the median change in $B/T$ is consistent with zero for
both morphological bins.  {\it Bottom Panel:} Same as above except for
the asymmetry parameter $R_A$.  Comparing the results for the original
and redshifted galaxies, we observe a slight decrease in the ability
to recover asymmetric features with decreased effective resolution
($0.02\pm0.01$).
\label{fig5}}
\end{figure}

\begin{figure}
\plotone{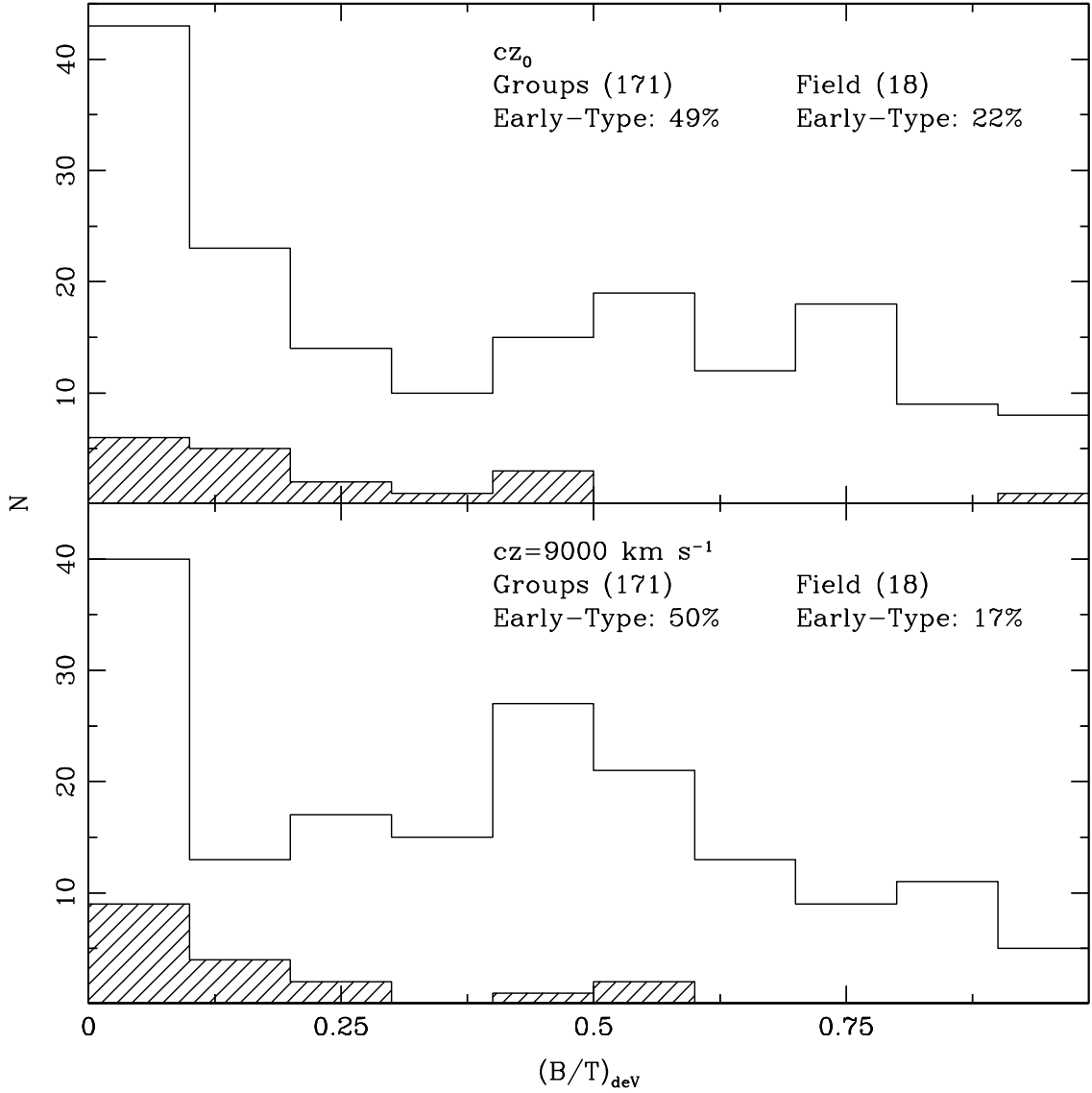}
\caption[fig6.eps]{Histogram of $B/T$ measured with GIM2D
using a 2D profile with a de Vaucouleurs bulge and exponential disk.
The bin size is 0.1.  The upper panels correspond to $B/T$ measured at
the observed redshift and the lower panels to $B/T$ measured at
$cz=9000$\kms.  The groups are the open bins and the field the
hatched bins.  A K-S test comparing the group and field distributions
at their observed redshifts and at $cz=9000$\kms~rule out a common
parent distribution with $>95$\% certainty in both cases.
\label{fig6}}
\end{figure}

\begin{figure}
\plotone{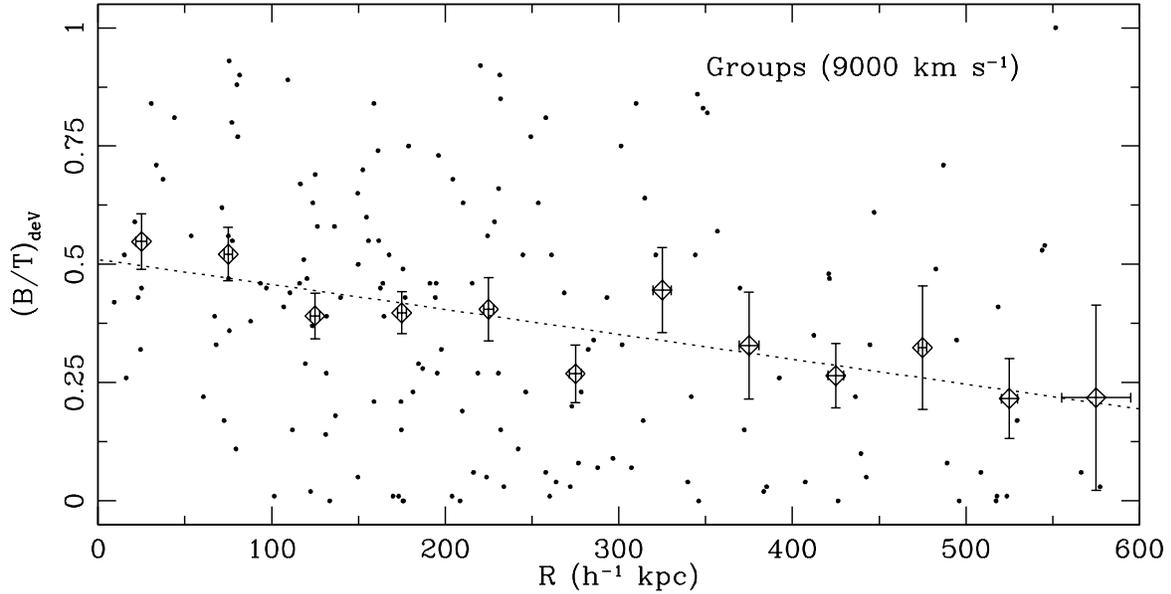}
\caption{$B/T$ as function of distance from group 
center for all group galaxies (171) at $cz=9000$\kms~and within
$R<0.5 h^{-1}$ Mpc, approximately the group virial radius.  The open
diamonds represent the average $B/T$ for each 50\hi kpc bin.  The
errorbars correspond to the standard error of the mean in the $R$ and
$B/T$ distributions for each bin.  There is a significant trend of
decreasing $B/T$ with increasing radius ($>95$\% significant using the
Spearman rank test).  The dotted line is the least-squares fit to the
data.  Despite their small number of members, these X-ray detected,
poor galaxy groups display a morphology-radius ($\sim$ density)
relationship similar to that of galaxy clusters.
\label{fig7}}
\end{figure}

\begin{figure}
\plotone{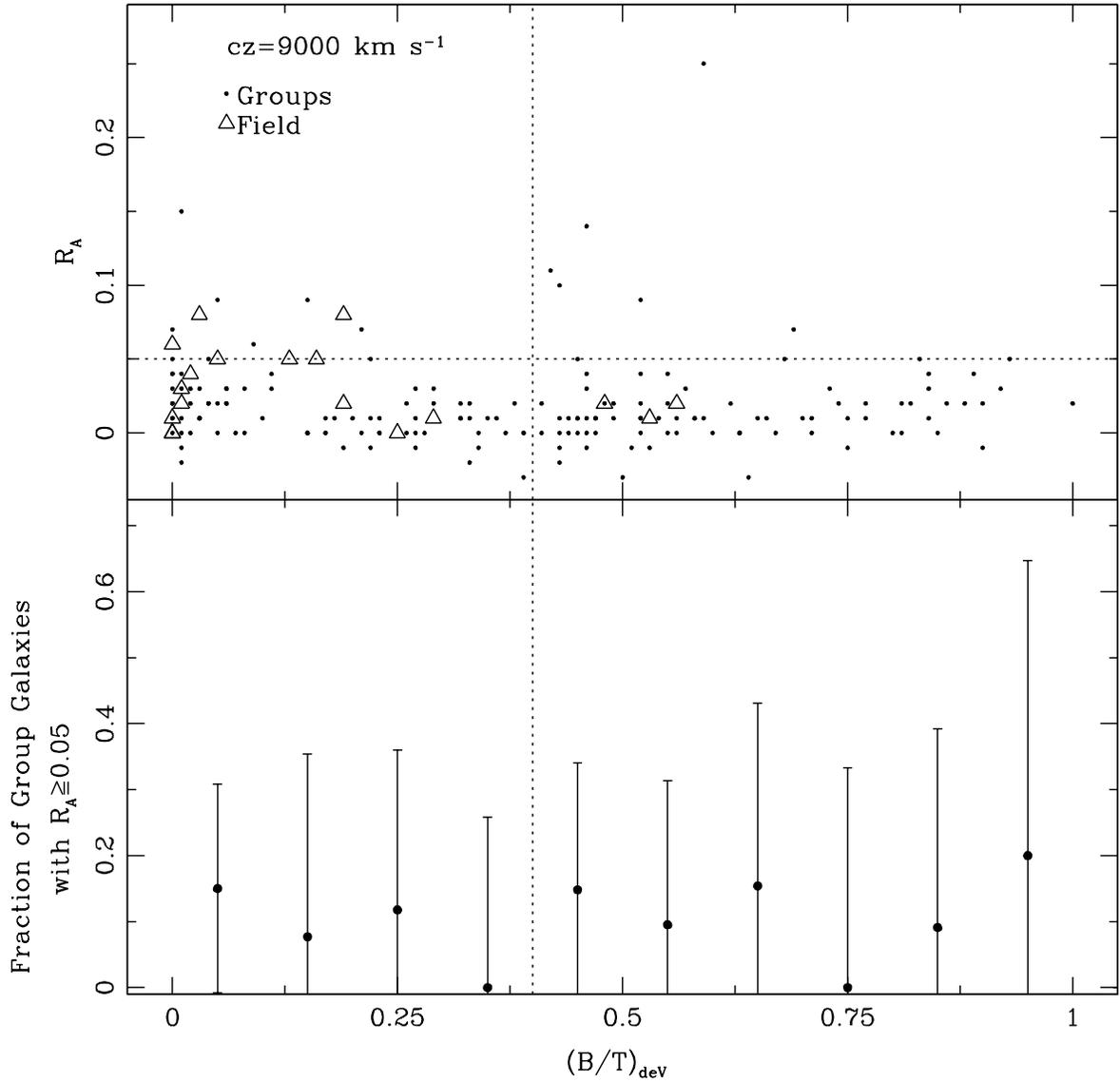}
\caption{Distribution of the asymmetry parameter
$R_A$ with respect to $B/T$ for the group and field populations at
$cz=9000$\kms.  {\it Top Panel:} $R_A$ as a function of $B/T$.  Group
members are the filled circles and the field are the open triangles.
The dotted lines represent $R_A=0.05$ and $B/T=0.4$, the defined lower
limits for asymmetric and bulge-dominated galaxies respectively.  {\it
Bottom Panel:} Fraction of group galaxies with significant galaxy
asymmetry ($R_A\geq0.05$) versus $B/T$; $1\sigma$ errorbars are shown
for each bin.  There is no significant correlation of $B/T$ with
either $R_A$ or the fraction of galaxies with high $R_A$.
\label{fig8}}
\end{figure}

\begin{figure}
\plotone{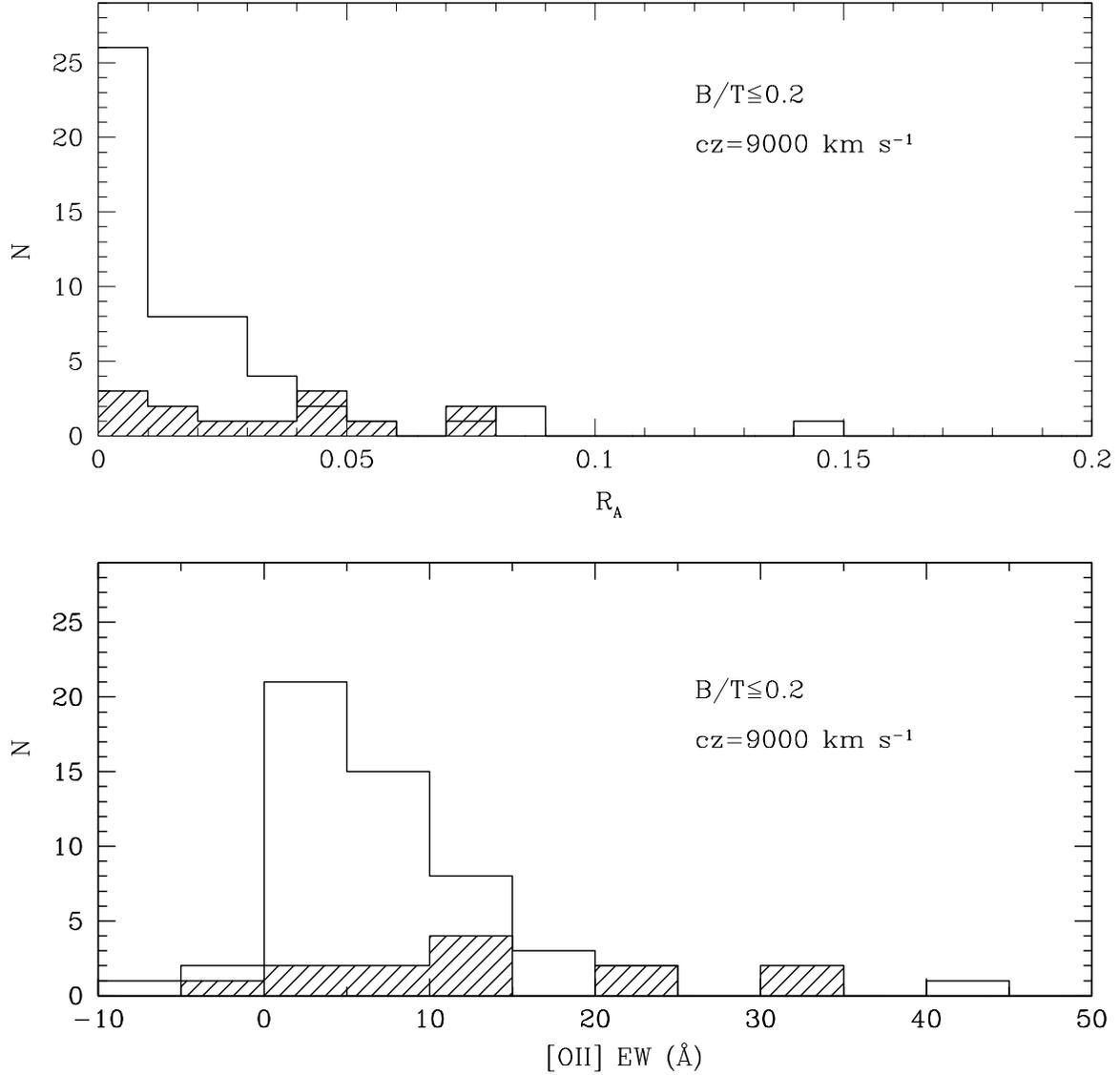}
\caption{{\it Top Panel:} $R_A$ distribution for
the most disk-dominated ($B/T<0.2$) group (open bins) and field
(hatched bins) galaxies.  {\it Bottom Panel:} [OII] EW distribution
for the same group and field galaxies.  In both panels, a K-S test
rules out a common distribution between group and field galaxies with
$>95$\% certainty.
\label{fig9}}
\end{figure}

\begin{figure}
\plotone{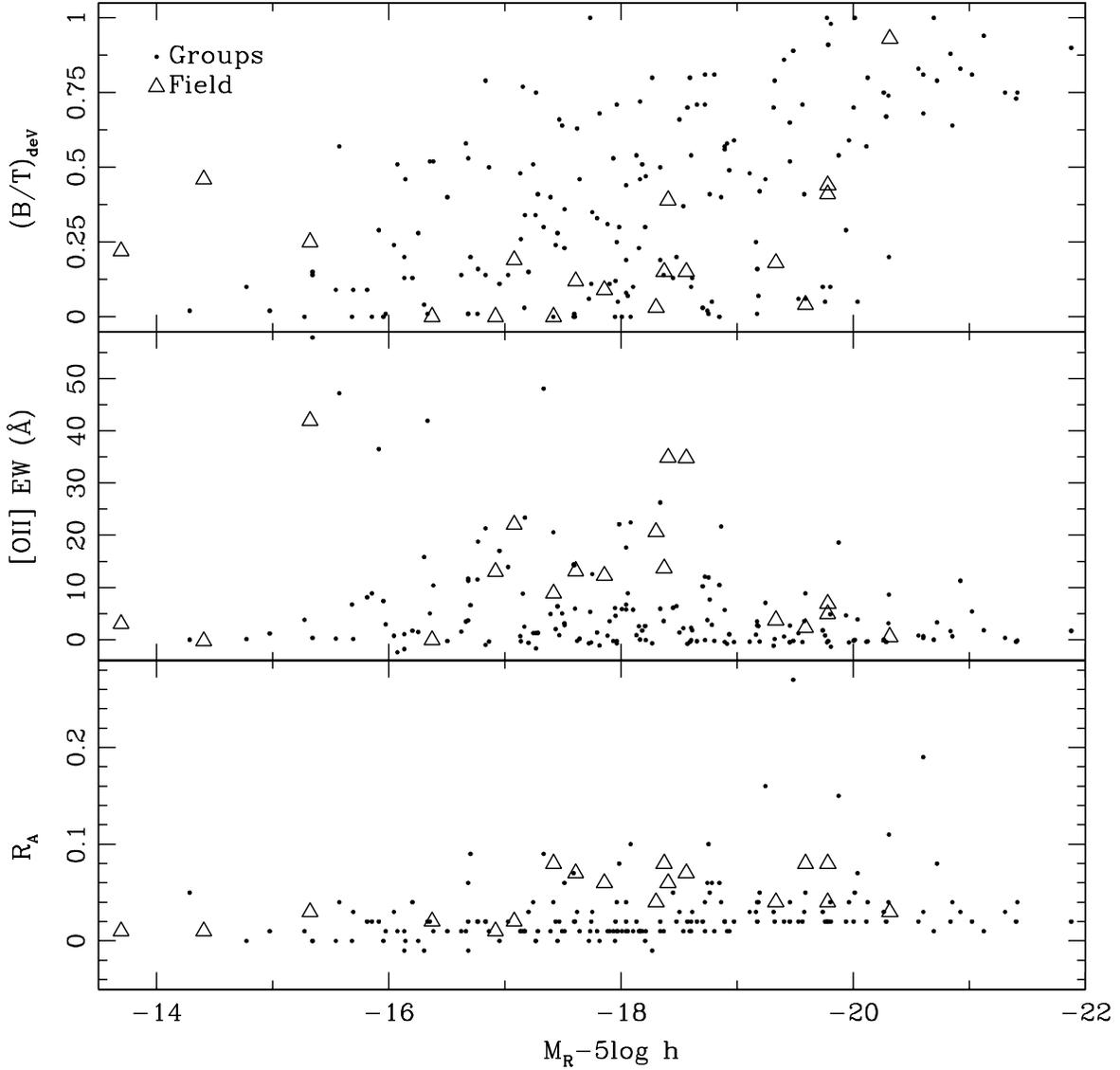}
\caption{Distribution of $B/T$, [OII] equivalent
width, and asymmetry $R_A$ versus absolute magnitude ($M_R$) for group
and field galaxies shown in the top, middle, and bottom panel,
respectively.  Despite having similar $M_R$ ranges, the group and
field galaxy populations differ in their $B/T$, [OII] EW, and $R_A$
distributions.
\label{fig10}}
\end{figure}

\begin{figure}
\plotone{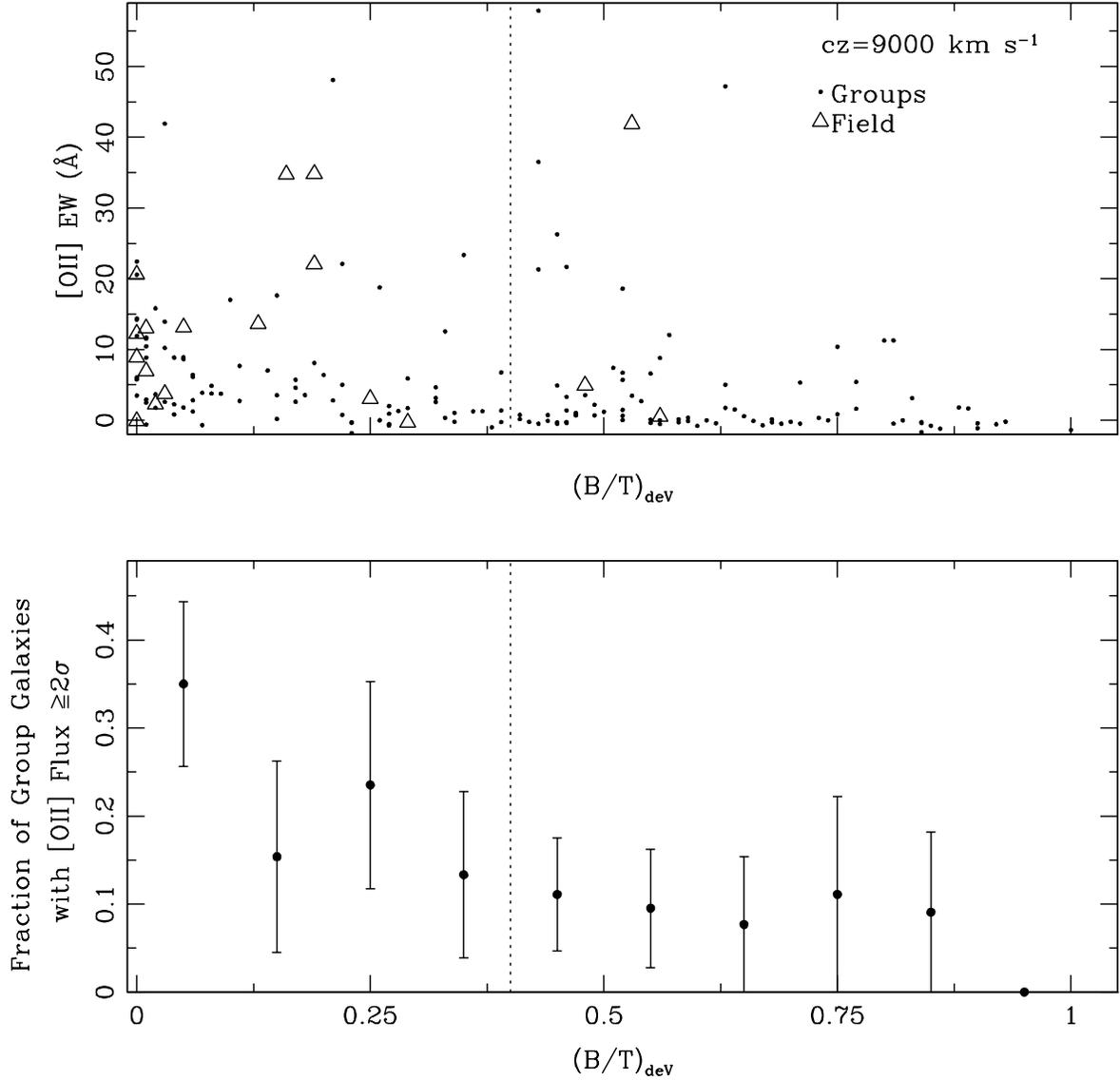}
\caption{Distribution of [OII] emission with respect to
$B/T$ for the groups and the field at $cz=9000$\kms.  {\it Top Panel:}
The [OII] equivalent width versus $B/T$.  The dotted line represents
$B/T=0.4$, the defined lower limit for bulge-dominated galaxies.  {\it
Bottom Panel:} Fraction of group galaxies with [OII] {\it flux}
greater than $2\sigma$ versus $B/T$, where $\sigma$ is the error in
the flux; 68\% confidence errorbars are shown for each bin.  The
fraction of galaxies with significant [OII] flux drops by almost two
beyond $B/T\sim 0.3$.
\label{fig11}}
\end{figure}

\begin{figure}
\plotone{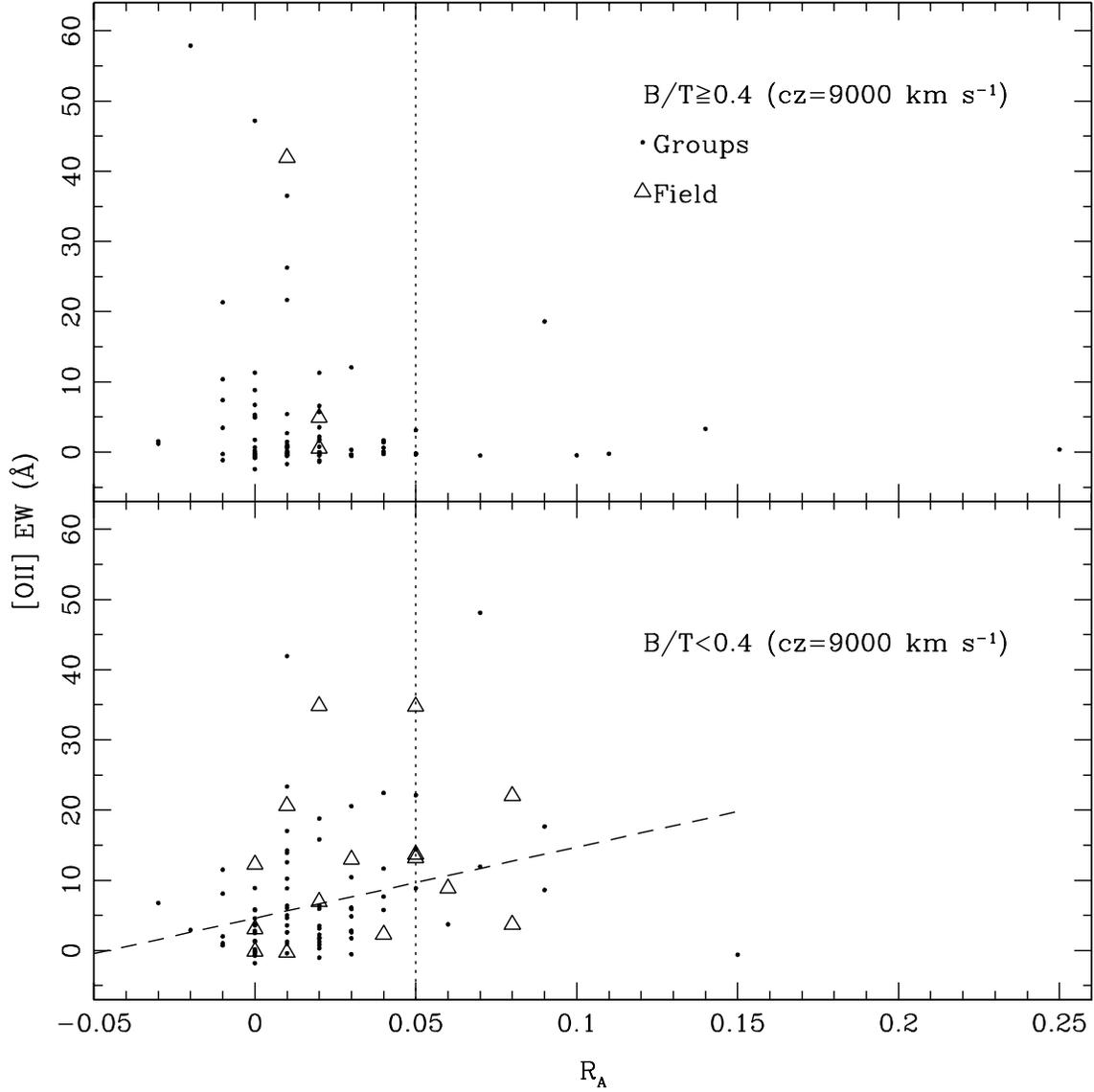}
\caption{Distribution of [OII] emission with respect to the
asymmetry parameter $R_A$ for the groups and the field.  {\it Top
Panel:} The [OII] equivalent width versus $R_A$ for early-type
galaxies ($B/T\geq0.4$).  The dotted line represents $R_A=0.05$, the
defined lower limit for asymmetric galaxies.  {\it Bottom Panel:} The
same except for late-type galaxies ($B/T<0.4$).  There is a
correlation between $R_A$ and [OII] EW for $B/T<0.4$ group galaxies;
the least-squares fit to only the group members is represented by the
dashed line in the $B/T<0.4$ panel.  The trend is $>95$\% significant
using the Spearman rank test.
\label{fig12}}
\end{figure}

\end{document}